\documentclass[acmlarge]{acmart}
\usepackage{fullpage}
\usepackage{url}
\usepackage{wrapfig}
\makeatletter
\def\@copyrightspace{\relax}
\makeatother
\usepackage{booktabs} % For formal tables
\usepackage{mdframed}
\usepackage{framed}
\usepackage{hyphenat}
\usepackage{algpseudocode}
\usepackage{amsmath,epsfig}
\usepackage[boxed,ruled,vlined,linesnumbered]{algorithm2e}
\usepackage{amsthm}
\usepackage{setspace}
\usepackage{enumitem}
\usepackage{ragged2e}
\usepackage{multirow}
\usepackage{hhline}
\usepackage{wrapfig}
\usepackage{amssymb}
\usepackage{caption}
\usepackage{subcaption}
\usepackage{graphicx}
\usepackage{amsmath,epsfig}
\usepackage{amsthm}

\newtheorem{attackgame}{ATTACK GAME}

\setcopyright{none}
\setcopyright{acmcopyright}
\setcopyright{acmlicensed}
\setcopyright{rightsretained}
\setcopyright{usgov}
\setcopyright{usgovmixed}
\setcopyright{cagov}
\setcopyright{cagovmixed}

\newcommand{\B}{\vspace*{-\smallskipamount}}
\newcommand{\BB}{\vspace*{-\medskipamount}}
\newcommand{\BBB}{\vspace*{-\bigskipamount}}
\newcommand{\xor}{\mathbin{\oplus}}

\newcommand{\aaa}{\mathcal{A}}

\newcommand{\aexp}{\mathrm{\mathbf{Exp}}_{\aaa}}

\acmJournal{TCPS}
\acmVolume{Unassigned}
\acmNumber{Unassigned}
\acmMonth{3}

\begin{document}

\title{\textsc{Canopy}: A Verifiable Privacy-Preserving Token Ring based Communication Protocol for Smart Homes}\titlenote{\textbf{A preliminary version of this paper was accepted in CODASPY 2019~\cite{DBLP:conf/codaspy/Panwar0WMV19}. \\ This version has been accepted in ACM Transactions on Cyber-Physical Systems.} The final published version of this paper may differ from this accepted version. \\ This material is based on research sponsored by DARPA under agreement number FA8750-16-2-0021. The U.S. Government is authorized to reproduce and distribute reprints for Governmental purposes notwithstanding any copyright notation thereon. The views and conclusions contained herein are those of the authors and should not be interpreted as necessarily representing the official policies or endorsements, either expressed or implied, of DARPA or the U.S. Government. This work is partially supported by NSF grants 1527536 and 1545071.}

\author{Nisha Panwar$^{1,2}$, Shantanu Sharma$^2$, Guoxi Wang$^2$, Sharad Mehrotra$^2$, and Nalini Venkatasubramanian$^2$}
\affiliation{%
	\institution{$^1$Augusta University, USA. $^2$University of California, Irvine, USA.}}

\begin{abstract}
%Advances in sensing, networking, and actuation technologies have resulted in the IoT wave that is expected to revolutionize all aspects of modern society. This paper focuses on the new challenges of privacy that arise in IoT in the context of smart homes. Specifically, the paper focuses on preventing the user's privacy via inferences through channel and in-home device activities. We propose a method for securely scheduling the devices while decoupling the device and channels activities. The proposed solution avoids any attacks that may reveal the coordinated schedule of the devices, and hence, also, assures that inferences that may compromise individual's privacy are not leaked due to device and channel level activities. Our experiments also validate the proposed approach, and consequently, an adversary cannot infer device and channel activities by just observing the network traffic.
%Advances in sensing, networking, and actuation technologies have resulted in the IoT wave that is expected to revolutionize all aspects of modern society.
This paper focuses on the new privacy challenges that arise in smart homes. Specifically, the paper focuses on inferring the user's activities -- which may, in turn, lead to the user's privacy --  via inferences through device activities and network traffic analysis. We develop techniques that are based on a cryptographically secure token circulation in a ring network consisting of smart home devices to prevent inferences from device activities, via \emph{device workflow}, \textit{i}.\textit{e}., inferences from a coordinated sequence of devices' actuation. The solution hides the device activity and corresponding channel activities, and thus, preserve the individual's activities. We also extend our solution to deal with a large number of devices and devices that produce large-sized data by implementing parallel rings. Our experiments also evaluate the performance in terms of communication overheads of the proposed approach and the obtained privacy.
\end{abstract}




\maketitle

%========================================SECTION=========================================

\section{Introduction}
\label{sec:introduction}

Smart home devices are quickly becoming an integral part of our routine life. These devices provide comfort/assisted-living, improve sustainability, reduce costs, and reduce carbon footprint. For example, a Belkin Wemo switch can automatically switch lights on/off and open/close window shades based on the sunlight and time of the day. Below, we consider an example to motivate how these smart home devices help us and then see why these devices are vulnerable to different attacks, by which they may jeopardise the user privacy.

%\smallskip
%\noindent\textbf{{Motivational example.}}

\subsection*{Motivational Example}

{We consider a use-case based on a smart assisted living setup for the elderly. According to statistics gathered by the U.S. Department of Health and Human Services, the United States is expecting to double its senior citizen (65+ age) population (currently $\approx$50 million) by 2060; today, about 28\% of all non-institutionalized seniors live alone~\cite{old-people,DBLP:journals/suscom/AlhassounUV19}.  To cater to this growing demographic, smart homes that enable independent living and aging, including homes and resort communities for people of 55+ age, are on the rise~\cite{55old}. Such homes are often  equipped with several IoT devices that create new capabilities and services to help occupants live, move, and age more comfortably, as well as, safely. Examples of such services (and the devices to enable them) include:
(\textit{i}) presence and movement services by motion sensors and activity detection devices;
(\textit{ii}) health monitoring services to keep physiological information that can be used to detect and predict potential ailments (\textit{e}.\textit{g}., gait analysis, early signs of neuro-degeneration) by health tracking devices worn by individuals;
(\textit{iii}) reminder services to users about appointments, meals, and medications by smart-speakers and voice-activated devices;
(\textit{iv}) cooking services by a variety of smart cooking device;
(\textit{v}) lightening services to improve visibility when light levels are low in selected areas of the home where movement is anticipated by smart lights;
(\textit{vi}) notification services to make medication accessible based on pre-established schedules, when doses are required or missed, by smart pillboxes; and
(\textit{vii}) fall detection services to detect when a fall has occurred, validate the need for help and alert caregivers as needed, by a variety of motion and fall detection sensors.}

{
The above-mentioned services can be viewed as a \emph{device workflow}, consisting of a coordinated sequence of devices' actuation. Based on the execution of device workflows, we classify them into three categories, as follows:}
\begin{enumerate}[noitemsep,nolistsep,leftmargin=0.1in]
  \item {\emph{Triggered/synchronized workflows}: These are device workflows that arise when events that cause actuation of a device and may result in subsequent actuation of other devices (see example below).}
  \item {\emph{Scheduled workflows}: These are device workflows that arise as a result of a planned/scheduled actuation of a set of devices and occur at a specific time intervals based on a schedule.}
  \item {\emph{Hybrid workflows}: These are device workflows that arise due to a possible combination of scheduled and triggered workflows.}
  \end{enumerate}

{The following scenario depicts the different types of workflows in the smart assisted living use-case above. Alice is an elderly lady, who lives alone and independently for much of the day; she is unable to walk fast and steadily, and has difficulty remembering when to take her medications. To assist her daily activities, her family members have created the following scheduled workflow: A smart pillbox will unlock and alert Alice at a predefined time in the morning for her first set of medications. Following this, a series of sensors (motion, activity sensing, and fall detection)  will be turned on to capture her expected Activities of Daily Living (ADLs). To ensure that her medication has been consumed, the smart pillbox will activate an on-board load sensor to detect changes in weight -- this will occur 15 minutes after the expected medication timing. Now, if the pillbox detects that the medication has not been taken from the box, appropriate reminders and alerts are issued to Alice (and potentially to caregivers).  Otherwise, the pillbox goes back into locked mode and repeats the above cycle later to ensure that Alice takes her evening medications. Meanwhile, gait analysis from smartpad pressure sensors on the floor indicates that Alice is more unsteady than usual. Smartpad sensors capture movement data at increased levels of granularity than normal and translate this into potential ADLs. Alice has a fall later that afternoon, the fall detection system is activated, acoustic and camera sensors will be triggered to capture images and converse with Alice to gather further details about her condition. The system confirms that Alice has likely had a fall and triggers alerts to caregivers and first responders. The above scenario illustrates a scheduled workflow for medication assistance (created by family members), as well as, a triggered workflow (caused due to Alice falling).}

{Consider another example of only a triggered workflow. Assume that Alice's front door is equipped with a smart door lock, a door camera, and a smart doorbell. When a guest rings the smart doorbell, it triggers a new event, \textit{i}.\textit{e}., turning on the door camera. The image captured by the camera is passed to Alice, who recognizes her friend at the door. She issues a voice command to unlock the smart door, (a similar use-case of the door opening is also considered in~\cite{carnemolla2018ageing}).} $\blacksquare$

\smallskip
{Observe that smart home IoT devices capture, store, share, or (depending upon the underlying computational architecture) outsource personal data. In addition, many of such IoT devices communicate wirelessly. Thus, one may be able to capture the network traffic and then learn device activities that may reveal
the device workflows. By revealing the device workflow, one may also learn the user activities, which may, in turn, reveal the user privacy (inferences about individual's habits, behavior, family dynamics, presence/absence of occupants in specific regions of the home, and scheduled activities).} Also, some of these devices do not provide an end-to-end secure architecture, by which an adversary can attack them. For example, McAfee Labs~\cite{guoxi1} found that the well-known WiFi-enabled Wemo Insight Smart Plug has critical security vulnerability due to Universal Plug and Play (UPnP) protocol library it uses, which, due to design flaws, enable attackers to execute remote codes on this smart plug. Note that this attack is not just limited to disturbing smart plug's normal operations such as shutting it down unexpectedly, but could also use the smart plug as an entry point for a larger attack in the network. Further,~\cite{guoxi1} showed the usage of a compromised WeMo switch as a middleman to launch attacks against a TCL smart TV.

%\begin{wrapfigure}{r}{0.45\textwidth}
\begin{figure*}[!t]%{r}{0.45\textwidth}
\BBB%\BBB\BBB
	\begin{center}
		\includegraphics[scale=0.55]{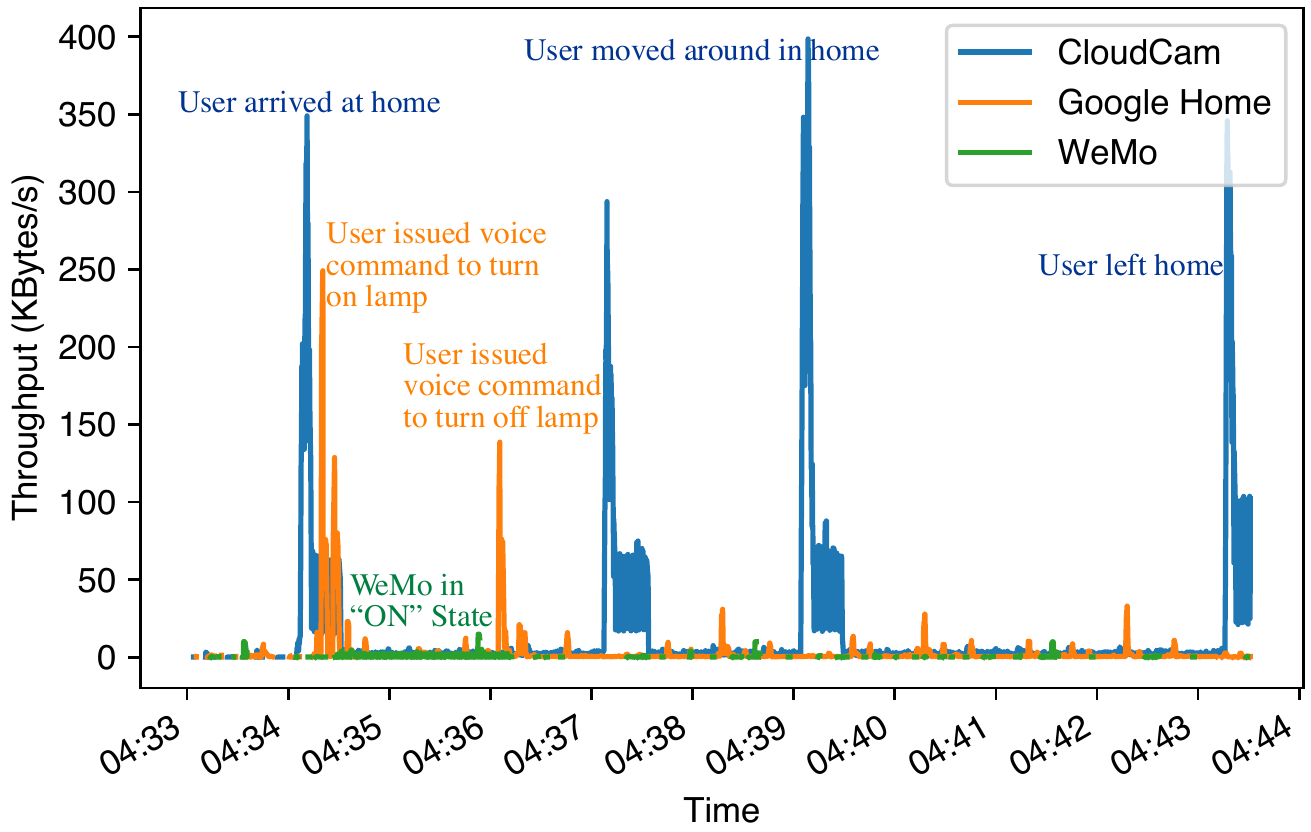}
	\end{center}
\BBB\BB
\caption{Channel activity for home devices}
%\begin{flushleft}

{{\scriptsize The figure above shows the channel activity for three home devices: CloudCam, Google Home, and Belkin WeMo. The CloudCam shows a peak in the channel activity (up to 400 KB/s traffic rate) as the user enters the home, moves inside the home or exits the home. The Google Home shows a peak in the channel activity (up to 250 KB/s traffic rate) whenever a user initiated a voice command for the light bulbs to turn on/off. Similarly, a bi-state WeMo switch peaks during the on state and creates a channel activity lesser than 20 KB/s.}}
	%\BBB
	\label{fig:peak}
\BBB
\end{figure*}

Moreover, challenges arise since privacy leakage can occur through direct data leakage, as well as, through inferences based on devices' actuation, interactions, and schedules, as our motivational example shows. While privacy challenges from data leakage can be prevented by encrypting device data and network payloads, inferences about device actuation and schedule are significantly more complex to hide due to leakage from network traffic patterns at the channel level, at the hub/router level, or at the cloud level. For example, Figure~\ref{fig:peak} shows the network traffic generated by three different devices, where each device generates a significantly distinct traffic pattern. The adversary, having access to the channel traffic, can deduce which device is activated, leading to potential inferences about the user's personal habits. It is important to note that such an inference -- that is independent of the actual network payload -- cannot be prevented by encryption. {It is important to mention that such inference attacks that identify/classify IoT devices using the network traffic patterns on a wireless network have been further studied in~\cite{pingpong,DBLP:journals/corr/abs-1808-02741,DBLP:conf/sigcomm/HamzaGS18,DBLP:conf/iotdi/OrtizCL19,DBLP:journals/tmc/SivanathanGLRWV19,DBLP:conf/sp/AlrawiLAM19}. For example, by analyzing the encrypted wireless traffic using machine learning techniques,~\cite{DBLP:journals/corr/abs-1808-02741} achieves above 90\% accuracy in identifying activities of IoT devices in a smart home.~\cite{DBLP:conf/sigcomm/HamzaGS18} also showed that one can analyze the network traffic and create a profile of a device based on communication patterns.} {Further, note that the adversary can trivially launch the sniffing attack, due to the following reasons: (\textit{i}) The devices required for sniffing attacks are easy to make at a low cost. For example, to capture WiFi packets, any commodity laptop with the proper software can enable us to launch a sniffing attack. Moreover, a pocket-size sniffer could be made using Raspberry Pi Zero W and a portable battery. (\textit{ii}) Launching such a passive attack may not even require the adversary to enter the target space. The adversary can place the wireless network sniffer close to the target space (around 10 meters away) is sufficient to capture enough packets to analyze the device network activity patterns.}\footnote{{\scriptsize {In our motivational experiment (see Figure~\ref{fig:peak}), besides placing the wireless network sniffer in the testbed room, we also tried experimenting with a sniffer deployed around three rooms away from the smart home testbed in the building. The results still show apparent network access patterns of the targeted IoT devices.}}}

Inferences from monitoring channel traffic can also arise due to the characteristics of the current network protocols. For instance, in
the widely used 802.11 WiFi protocol, while a message payload is encrypted in a password-protected network, the MAC addresses of both the sender and receiver are in cleartext. This prevents every device on the network to decrypt a message just to determine if the message is intended for the device. {The leakage of the sender/receiver MAC address -- coupled with several other facts such as the network traffic characteristics~\cite{DBLP:journals/tmc/SivanathanGLRWV19} and public manufacturer's information -- may leakage the identity of the device that, in turn, can lead to inferring the user's activities.\footnote{{\scriptsize {It is important to mention that based only on MAC address, it is not trivial to find the device~\cite{DBLP:journals/tmc/SivanathanGLRWV19}.}}} For example, MAC address of Amazon CloudCam security camera used in our experiment to transmit video footage over WiFi was F0-81-73-23-CC-75. The first 3 bytes of such a MAC address (\textit{i}.\textit{e}., F0-81-73) can be searched in the publicly accessible IEEE Organizationally Unique Identifier (OUI)~\cite{2guoxi} dataset to find the vendor related information (\textit{e}.\textit{g}., Amazon device). Furthermore, by monitoring the device's traffic patterns and the fact that Amazon only manufactures a limited range of devices (\textit{e}.\textit{g}., Kindle, CloudCam, and Echo), it is easier to infer the device type by merely overhearing the traffic.}

%\medskip
%\noindent
%{\textbf{Problem.}

\subsection*{Problem}
This paper deals with the problem of avoiding inference attacks on user activities that arise when the adversary is able to observe device workflows in a home network. The workflows can be identified through \emph{coupling between channel and device activities}. We focus on the scheduled workflows, since such workflows require us to deal with two crucial concepts that are subject to privacy violations: workflow (\textit{i}.\textit{e}., the specific order of device actuation) and workflow execution (\textit{i}.\textit{e}., duration in which the devices coordinate, and the resulting device actuations unfold). We deal with hiding both the workflows and their execution. The scheduled workflows are important to deal with, since they do not just allow the adversary in learning the user's past behavior, but also the future behavior, since this past behavior is reflective of their future activities.

{We develop solutions to address two specific types of threats arise in this context:
(\textit{i}) overhearing of channel activity that may lead to inferring device activity patterns, and
(\textit{ii}) accessing the device temporarily so as to analyze the state of workflow execution.
In the latter case, the adversary can read the sent/received messages or view the internal state of the device and use that information to predict user activities, such as presence/absence, arrival/departure, and localization. We develop solutions to deal with both types of adversaries. It is also important to mention that mechanisms to prevent leakages for scheduled workflows will also prevent leakage from triggered or hybrid workflows, as will be clear soon.}

%\smallskip
%\noindent\textbf{Contributions.}

\subsection*{Contribution}
Our contributions are twofold:
\begin{enumerate}[noitemsep,nolistsep,leftmargin=0.01in]
  \item A new architecture for in-home communication among the devices and the hub through passing a token that is carrying commands. The token passing communication model decouples the channel and device activities, thereby the devices interact with the hub, as well as, with other devices without revealing the communication pattern. In addition, this architecture is useful for secure data upload from devices to the hub, while also hiding the device footprints to reveal which device has generated the data.

  \item An owner-defined pre-scheduling mechanism for all devices that are connected with the hub in a pre-defined topology. The proposed approach uses a single message transmission for all devices, while ensuring that in-home communication remains peakless (\textit{i}.\textit{e}., identical regardless of the type of devices and their activities). The scheduling mechanism is secure against a computationally unbounded adversary and, also, verifies the delay between each device actuation.
\end{enumerate}

%\smallskip
%\noindent{\textit{Practicality of the proposed technique.}

\subsection*{Practicality of The Proposed Technique}
{The proposed technique does not require to change the physical smart home devices. Instead, our technique is compatible with modern smart home IoT systems and only requires software updates on devices and the gateway (note that such software updates are common in several smart IoT devices, such as surveillance cameras, TVs, voice controlled devices (Amazon Echo), and smart plugs). In particular, the hub's functionalities could be either integrated into the smart home gateways or implemented on an existing computing unit in a home; for example, iPad/Apple TV in Apple's Homekit system, Alexa Echo Show, and Google Home Hub. The hub functionalities only require a moderate level of computing capability. As we will see in the experiment, we use a NUC with 5th gen Core ultra-low-power (U) CPU as the hub to handle around 70 simulated devices in a smart home setting. The used hub in our experiments had a similar capability to Apple TV in Apple's Homekit system, Alexa Echo Show, and Google Home Hub. In addition, we do not need any specific hardware or computing capability to solve the puzzle and execute the command. Our experiments used Raspberry Pi gen3 with a 1.2GHz ARM Cortex A-53 CPU as an IoT device. We selected this platform since it is commonly used in many IoT testbeds experiments~\cite{DBLP:journals/tecs/ZhouSY19,DBLP:journals/tr/SiboniSMBMBSE19,DBLP:conf/smartcomp/UddinNBWZHACSHD16}. However, existing IoT devices (sensors/actuators) generally use ARM Cortex M level processors (\textit{e}.\textit{g}., Texas Instrument Sensor Tags) that are less powerful than the Cortex A-series CPUs. Nevertheless, the proposed techniques can be implemented in the existing devices for the following two reasons: First, our  implementation is in Python scripting language for development simplicity. The same functions could be implemented using efficient languages such as C, which is also used for programming over ARM Cortex M processors~\cite{zhu2017embedded}. Second, the RSA puzzles that are used to enable an artificial verifiable delay could be carefully tuned to adapt to the computing capability of the underlying IoT device.} The proposed approach is also highly responsive in terms of the execution of the user's command. Particularly, the user commands can be executed in at most 250 milliseconds in a ring network of 75 devices, or at most 90 milliseconds in a network of three parallel rings (Experiment 4).

\subsection*{Outline}
The paper proceeds as follows: Section~\ref{sec:Preliminaries} provides the system setting, the adversarial model, security goals, and design requirements. Section~\ref{sec:decoup} provides our proposed scheduling algorithm for home networks. Section~\ref{sec:analysis} provides proofs of security and privacy. Section~\ref{sec:extend} provides improvements over basic scheme construction. Finally, Section~\ref{sec:Experimental Evaluation} provides an experimental evaluation of the proposed scheme. All notations are given in Table~\ref{table:notations}.

\section{Preliminaries}
\label{sec:Preliminaries}
This section presents the system model, the adversarial model, inference attacks on the user privacy, an overview of our proposed approach to prevent inference attacks, design requirements, and building blocks of the proposed algorithms.

\begin{table}[t]
\BBB\BBB\BBB
	\begin{center}
		%\begin{spacing}{0.45}
		\begin{tabular}{l l l l l l}
			\hline
			%            \multicolumn{10}{|c|}{Country List} \\
			%            \hline
			{\scriptsize Notations} & {\scriptsize Meaning} & {\scriptsize Notations} & {\scriptsize Meaning} & {\scriptsize Notations} & {\scriptsize Meaning}\\ \hline
			
			{\scriptsize $O$} & {\scriptsize A homeowner} & {\scriptsize $H$} & {\scriptsize Hub} & {\scriptsize $\mathcal{R}$} & {\scriptsize partial order} \\
			
			{\scriptsize $O_{\mathit{id}}$} & {\scriptsize Owner's identity} & {\scriptsize ${H}_{\mathit{id}}$} & {\scriptsize Hub identity} & {\scriptsize $\mathcal{E}^\prime$} & {\scriptsize partially ordered set} \\ %\hline
			
			{\scriptsize $D$} & {\scriptsize Device} & {\scriptsize $D_{id}$} & {\scriptsize Device identity} & {\scriptsize $\varmathbb{P}$} & {\scriptsize device profile}\\ %\hline
			
			{\scriptsize $O_{\mathit{PK}}$} & {\scriptsize Owner public key} & {\scriptsize $O_{\mathit{SK}}$} & {\scriptsize Owner secret key}  & {\scriptsize $r_i$} & {\scriptsize $i$th ring instance} \\ %\hline
			
			{\scriptsize $H_{\mathit{PK}}$} & {\scriptsize hub public key} & {\scriptsize $H_{\mathit{SK}}$} & {\scriptsize hub secret key} & {\scriptsize $\rho$} & {\scriptsize average load or number of commands} \\ %\hline
			
			{\scriptsize ${D}_{\mathit{PK}}$} & {\scriptsize device public key} & {\scriptsize ${D}_{\mathit{SK}}$} & {\scriptsize device secret key}  & {\scriptsize $\gamma$} & {\scriptsize peer device dependency} \\ %\hline
			
			{\scriptsize $c_l$} & {\scriptsize partially ordered $l$ commands} & {\scriptsize $m^i$} & {\scriptsize puzzle message for $i$th device} & {\scriptsize $\mathcal{A}$} & {\scriptsize adversary}\\ %\hline
			
			{\scriptsize $n$} & {\scriptsize modulus} & {\scriptsize $a$} & {\scriptsize random chosen integer}  & {\scriptsize $\hat{a}$} & {\scriptsize malicious commands} \\ %\hline
			
			{\scriptsize $\hat{t}$} & {\scriptsize time complexity of puzzle} & {\scriptsize $t^{\prime}$} & {\scriptsize time to decrypt command} & {\scriptsize $\mathcal{T}$} & {\scriptsize token} \\ %\hline
			
			{\scriptsize $S$} & {\scriptsize capacity of puzzle solver} & {\scriptsize $S^{\prime}$} & {\scriptsize enhanced capacity}  & {\scriptsize $\mathcal{H}$} & {\scriptsize One-way hash function}  \\ %\hline
			
			{\scriptsize $\mathit{Sign}(O_{\mathit{SK}})$} & {\scriptsize Signature using secret-key $O_{\mathit{SK}}$} & {\scriptsize $N$} & {\scriptsize number of devices} & {\scriptsize $\mathit{Data \: field}$} & {\scriptsize data upload field} \\ %\hline
			
			{\scriptsize $E_k$} & {\scriptsize encrypted key $k$} & {\scriptsize $E_z$} & {\scriptsize encrypted command $z$} & {\scriptsize $b_{\mathit{toggle}}$} & {\scriptsize toggle bit string field}\\ %\hline
			
			{\scriptsize $k_s$} & {\scriptsize static key} & {\scriptsize $^nP_n$} & {\scriptsize $n$ permutations} & {\scriptsize $t_{\mathit{diff}}$} & {\scriptsize allowed clock drift time} \\ %\hline
			
			{\scriptsize $\phi{(n)}$} & {\scriptsize Euler's totient on $n$} & {\scriptsize $b_o$} & {\scriptsize overwritten data bits} & {\scriptsize $t_{\mathit{com}}$} & {\scriptsize total computation time} \\ %\hline
			
			{\scriptsize $b_g$} & {\scriptsize device-generated data bits} & {\scriptsize $b_r$} & {\scriptsize random data bits} & {\scriptsize $t^{\mathcal{A}}_{\mathit{com}}$} & {\scriptsize puzzle computation time by adversary} \\ %\hline
			
			{\scriptsize $p$} & {\scriptsize large prime number} & {\scriptsize $q$} & {\scriptsize second large prime number} & {\scriptsize $t^H_{\mathit{end}}$} & {\scriptsize token round ending time at hub} \\ %\hline
			
			{\scriptsize $t_{\mathit{val}}$} & {\scriptsize command validity time} & {\scriptsize $t_{\mathit{cur}}$} & {\scriptsize current time}  & {\scriptsize $\epsilon$} & {\scriptsize negligibly small value}\\ %\hline
			
			{\scriptsize $t^i_{\mathit{rcv}}$} & {\scriptsize token receiving time} & {\scriptsize $t^i_{\mathit{fwd}}$} & {\scriptsize token forwarding time} & 	
			{\scriptsize $t^H_{\mathit{beg}}$} & {\scriptsize token round beginning time at hub} \\   %\hline

			 \hline
		\end{tabular}%\F
		\caption{Notations}
		\label{table:notations}
\BBB\BBB\BBB
		%\end{spacing}
	\end{center}
\end{table}

\subsection{The Model}
\label{subsec:mode}

\noindent\textbf{Network assumptions.} We consider a collection of $N$ ($D^1,D^2,\ldots,D^N$) heterogenous smart home devices that provide different functionalities to the homeowner $O$. Each device $D^i$ has a unique identity, denoted by $D^i_{\mathit{id}}$.
We consider that devices may have heterogeneous hardware/software underneath and may be located on different spatial dimensions (devices that are not in the line-of-sight). The devices can move within the home space, and thus, may have a different set of peer devices at different time intervals. Further, we assume that each device possesses a read-only hardware clock, and due to the ad-hoc nature of devices, we assume a clock drift within the bound $t_{\mathit{diff}}$, thereby two clocks cannot differ beyond $t_{\mathit{diff}}$ amount of time.

The owner initializes the devices and a controlling hub, $H$, using proper security mechanisms. We will list our assumptions about the underlying security mechanism below. In our model, the network is configured as a ring topology, which poses an ordering among devices, unlike the model that the current smart home devices use, where the owner communicates directly to the desired device via the hub. This ring topology could be built directly among devices and hub, if the communication protocols they use have P2P communication capability, like Zig-Bee, BLE, and WiFi. Alternatively, it could be built as an overlay on top of a star topology network, like WiFi infrastructure mode. In this case, if a device tries to forward the message to the next device in the proposed ring topology, then it needs to first send the message to the hub, and the hub, then, directly forwards the message to its next device. Note that we do not discuss a failure-resilient ring topology and existing fault-tolerant schemes can be leveraged here.

The owner sends workflows to the home devices through the hub. After receiving a workflow, the device gets actuated, stores its corresponding command, and forwards the workflow received from the previous device to the next device in the topology. After executing the command, the device may generate the data (for example, Nest camera starts recording and sends data whenever motion is detected). We use the ring topology to send this data to the hub that may be stored at the hub or may be transmitted to the cloud.\footnote{{\scriptsize Recall that we are not dealing with how the data will be transmitted from the hub
to the cloud without revealing anything. Our solution hides any activity within the home, \textit{i}.\textit{e}., how the data will be transmitted by the device to the hub without any privacy violations.}} In this paper, we use the words `command execution' and `workflow execution' interchangeably.

\noindent\emph{Token.} In our ring topology, we circulate a token that has three fields: (\textit{i}) command field, which carries a computational puzzle and the workflow, (\textit{ii}) data field, which carries the data generated by devices to deliver to the hub, and (\textit{iii}) toggle bit string, which is used to indicate which device has generated the data to the hub. Details of the token are given in Section 3.

\medskip
\noindent\textbf{Security assumptions.} Each device, hub, and homeowner possess the corresponding signing key-pair, \textit{i}.\textit{e}., ($D^i_{\mathit{SK}},D^i_{\mathit{PK}}$), ($H_{\mathit{SK}},H_{\mathit{PK}}$), and ($O_{\mathit{SK}},O_{\mathit{PK}}$), respectively. We do not assume an arbitrary behavior from the owner or the hub. The \emph{homeowner and hub} mutually verify the identities of each other through digital signatures in order to build trust between them. Therefore, the hub and the owner trust each other, and an adversary cannot compromise either the homeowner or the hub. Further, \emph{the hub and devices} build their own trust that is also based on the knowledge of a certified public-private key-pair of home devices.

\subsection{Adversarial Model}
\label{subsec:Adversarial Model}
The devices execute user-defined commands or workflows, as mentioned previously. {The adversary wishes to learn the commands or workflows and the execution of workflows. Thus, the adversary could perform the wireless message (or packet) sniffing to get access to the secure (encrypted) wireless messages flowing among the devices and the hub/homeowner. Based on the encrypted network messages, the adversary wishes to infer the user activities, which may, in turn, reveal the user's privacy. Since these messages are in the local wireless network, this access is pre-network address translation (NAT). This type of adversarial attack, \textit{i}.\textit{e}., wireless message sniffing, is similar to the adversarial model considered in~\cite{watson,thir}.} Further, {we assume that the adversary knows the number of smart devices in the home. The reason of this: the wireless protocols (\textit{e}.\textit{g}., Wi-Fi, Bluetooth, Zigbee) that are commonly used in the smart home settings, MAC addresses of the sender and the receiver are in cleartext form in the message (or packet) flowing on the wireless channel. Thus, the adversary may easily count the unique number of MAC addresses of devices in a home.}\footnote{{\scriptsize {Recently, the industry~\cite{stites2014user} became aware of the privacy risk of using cleartext MAC addresses. Thus, MAC address randomization methods are proposed and implemented on the majority of modern mobile devices, \textit{e}.\textit{g}., iOS devices, Android phones, Windows 10 devices, and some Linux devices~\cite{DBLP:conf/ccs/VanhoefMCCP16}. However, this mechanism does not prevent the adversary from knowing the number of home devices nor prevent wireless traffic pattern analysis. The reason is that most of the current MAC address randomization techniques, especially those on iOS and Android devices, are performed only in the active scanning phase, where the device broadcasts the probe request message, before the connection establishment. When the devices establish a connection with an access point, they still use static MAC addresses~\cite{DBLP:journals/popets/MartinMDFBRRB17}. Also, such randomization techniques are mostly implemented on phones/PCs and rarely seen on resource-constrained IoT devices.}}} Based on this information, the adversary aims to learn: (\textit{i}) the device activities, and (\textit{ii}) coupling between the channel and the device activities. However, the adversary cannot inject any fabricated messages over the channel to assess the state of the devices.

{While we focus on a passive adversary listening to the network traffic, we also consider a stronger adversarial model, where the adversary may gain short-term physical access\footnote{{\scriptsize For example, insurance inspection authorities/care-takers/home cleaners have to visit the home for a periodic inspection in the absence of the homeowner. In this case, the inspection might be related to any leakage detection, maintenance issues, insurance issues, etc. This short visit to the home for an inspection may allow them to monitor and check home devices as well. However, our proposed solution prevents such an adversary to analyze the current state of the devices or the activity pattern of the devices in near future.}} to the device, and hence, may retrieve the device state or messages.} The objective behind gaining short-term access to any device is to predict the future workflows of devices to infer the user activity. {In particular, under such an adversary having short-term access to the device}, our objective is to prevent the adversary to know (\textit{i}) which device has received the messages at which time, (\textit{ii}) when a device would execute the command, and (\textit{iii}) which devices have executed the command at which time.\footnote{{\scriptsize The command execution could enable the device to produce visible or auditory cues such as blinking lights or machine being activated which, in turn, may leak the state of the device. Such inferences from physical cues are outside the scope of the paper. We assume that the adversary does not have access to such device data.}}

\medskip
\noindent\textbf{Adversarial view and inference attacks.} When the user wishes to execute any command at a smart home device, an adversary knows which device received the message from the user at what time due to the network traffic generated by the user. Note that this information is revealed, because in network protocols such as 802.11 MAC addresses or the device identifiers are transmitted in cleartext, and only the payload is encrypted, as mentioned in Section~\ref{sec:introduction}. Further, the device may also produce some data in response to the requested message, and this also reveals to the adversary which device has generated the data at what time. We refer such information as the adversarial view, denoted by $\mathit{AV}$:
$\mathit{AV} = \mathit{In}_c \cup \mathit{Op}_d$, where $\mathit{In}_c$ refers to the command given to the device at some time and $\mathit{Op}_d$ refers to the data generated by the device after executing the command. %at some time.

\begin{table}[h]
\BBB
	%\scriptsize
	\centering
	\begin{tabular}{|l|l|l|}
		\hline
		Users command & \multicolumn{2}{|c|}{Adversarial view}           \\ \hline
		~                          & $In_c$ & $Op_d$ \\ \hline
		For $D^1$ & $\mathit{E(c_1)}, D^1, t_1$        & No        \\ \hline
		For $D^2$ & $\mathit{E(c_2)}, D^2, t_2$        & $\mathit{E(d_2)}, D^2, t_3$        \\ \hline
	\end{tabular}
	\caption{Adversarial view due to observing channel activities}
	\label{tab:Adversarial view}
\BBB\BBB\BB
\end{table}

For example, consider that there are two devices, say $D^1$ and $D^2$, in the home. In Table~\ref{tab:Adversarial view}, the first row shows that the user transmits a command to the device $D^1$. Though the device $D^1$ receives this encrypted command, denoted by $E(c_1)$, the adversary knows that a command is received at time $t_1$ by the device $D^1$ and the device $D^1$ has not generated any data in response the command. The second row shows that the adversary knows the device $D^2$ receives an encrypted command $E(c_2)$ at time $t_2$ and generated encrypted data $E(d_2)$ at time $t_3$. Hence, simply based on the above characteristic of the arrival of a message and generation of data, the adversary can determine which device was actuated.

\subsection{Preventing Inference Attacks: An Overview of Our Approach}
\label{subsec:Preventing Inference Attacks}
In order to prevent inference attacks, we develop an approach that decouples the device and channel activities. In short, the approach provides an ability to pre-schedule a set of commands for home devices, where the homeowner defines: (\textit{a}) what should be the workflow/schedule of home devices, and (\textit{b}) when should the devices execute a workflow. Informally, the proposed approach works as follows:

\begin{enumerate}[noitemsep,nolistsep,leftmargin=0.01in]
\item The owner invokes the hub by sending an encrypted schedule or workflow of the devices. Here, the hub authenticates the owner to validate the encrypted schedule. This step provides a guarantee that no channel spoofing or message re-transmission have occurred.
	
\item After a successful authentication phase, the hub creates a token ($\mathcal{T}$) to circulate the schedule to be executed by devices. This token rotates continuously in the topology. Note that in the token, device identifiers or MAC addresses are also encrypted. Further note that whenever the user wishes to transmit a schedule to devices, the immediate next round of token originated by the hub carries the encrypted schedule, and after that, the encrypted schedule is replaced by a random message to maintain the token size constant.
	
	\item {On receiving the token, each device retrieves the encrypted schedule that carries device-specific commands. Each device must complete a computation task before executing the original command. This computation task is referred to as a puzzle throughout the paper. Then, the device must check the puzzle validity time $t_{\mathit{val}}$, by using the current time $t_{\mathit{cur}}$, and the allowed clock difference $t_{\mathit{diff}}$. If the timer has not expired yet, the device decrypts the message, executes the puzzle to retrieve the real command to be executed.}
	
	\item As soon as a device finishes the command execution, it may generate data to be uploaded at the hub (as mentioned in Section~\ref{subsec:mode}). Now to hide which device has generated the data, each device follows a request-based approach, where a device $D^i$ flips the $i^{\mathit{th}}$ bit inside one of the fields of the token to indicate the need to upload freshly generated data in an anonymous manner. As a result of $i^{\mathit{th}}$ bit flipping, the hub knows that the device $D^i$ has requested the data upload, and in the next token cycle, the device $D^i$ appends the data inside a dedicated field of the token.
\end{enumerate}
Note that since a constant size token flows regularly in the topology of the home devices, an adversary observing the network traffic cannot distinguish which device has received a command at which time (due to step 2) and which devices have generated the data (due to step 4). Hence, based on this approach, the adversarial view for each round of token has the same information, which prevents inference attacks based on the devices and channel activities.

\noindent\textbf{Aside.} {As the token rotates continuously in the ring network, it may increase the delay to complete a single round of token circulation. Since several devices work on 802.11 that supports data transmission speed of 54Mbps to more than 500Mbps. If we have devices that need data transmission speed of at most 4Mbps (for example, video streaming speed suggested by Netflix), then our approach could support at most 10 such devices, when having 802.11 supporting 54Mbps. Since it is not often that all highly data-hungry devices work in parallel, we can support a good number of devices at a time, while using 802.11 based communication.}

\subsection{Security Goals}
This section describes the security properties for preventing any inferences about the workflow and their executions from the adversary. Let us assume that an adversary knows some auxiliary information about the devices and the topology such as the number of devices and the types of devices. However, this auxiliary information does not increase the probabilistic advantage that an adversary gains over any instance of the protocol. In particular, an adversary cannot reveal the workflow or the execution time of the workflow, \textit{i}.\textit{e}., which devices execute the command or when does a device execute the command. The probabilistic advantage of an adversary, denoted by $\mathit{Adv(\mathcal{A})}$, is derived through the security properties given below.

\smallskip\noindent\textbf{Authentication} is required during the workflow release from the homeowner to the hub. This would require a mutual authentication between the (mobile device held by the) homeowner (to dispatch the workflow) and the hub (to circulate the workflow anonymously). Note that establishing the shared secret between the homeowner and the hub is a one-time process, which is carried each time the homeowner invokes a new workflow. Here, the homeowner produces a signature, say $\mathit{Sign}$, on the ordered commands ($c_l$) in any workflow by using its secret-key $O_{SK}$. Thus, the hub must reject any other messages signed by a different key, say $O_{{SK}^\prime}$.

\centerline{$Pr[(O_{SK},c_l)\rightarrow \mathit{Sign}]\geq 1-\epsilon$}

Note that $\epsilon$ is negligibly small and an adversary cannot produce a verifiable signature $\mathit{Sign}$ on $c_l$ by using $O_{{SK}^\prime}$ instead of $O_{SK}$.

\smallskip\noindent\textbf{Anonymity} is required during the consistent circulation of the encrypted commands such that (\textit{i}) no channel activity can be mapped to a device activity, and (\textit{ii}) no inference on device activity can be mapped to the device-generated data, \textit{i}.\textit{e}., which device is sending data at a specific time. As shown below, the probability of distinguishing two different tokens ($\mathcal{T},\mathcal{T}^\prime$) each carrying different messages ($m^i,m^j$) for different devices ($i,j$) is negligible.

\centerline{$Pr[\mathcal{T}(m^i)]-Pr[\mathcal{T}^\prime(m^j)]< \epsilon$}

Similarly, the probability of distinguishing a token $\mathcal{T}$ carrying the random data $b_r$ or carrying the overwritten data $b_o$, is negligibly small. Therefore, a token carrying the random data inside the data field and another token carrying the overwritten data (after data generation) inside the data field are indistinguishable, hence, solely based on the token data field no inferences can be derived.

\centerline{$Pr[\mathcal{T}(b_r)]-Pr[\mathcal{T}(b_o)]< \epsilon$}

\smallskip\noindent\textbf{Verifiable delay} An adversary cannot infer the information about the device execution ahead of time. Let $\hat{t}_i$ be the time a device would execute even when the adversary has temporary access to the device.

\centerline{$Pr[t^{\mathcal{A}}_{\mathit{com}}|\mathit{state}] \approxeq Pr[t^{\mathcal{A}}_{\mathit{com}}]$}

The probability of an adversary finishing the computation task earlier, when it gains temporary access to the device state, is approximately same as when the adversary does not have access to the device state. In addition, an adversary cannot outpace a device that requires $\hat{t}^i$ time to complete the computation task, \textit{i}.\textit{e}., $Pr[t^{\mathcal{A}}_{\mathit{com}}<\hat{t}_i]< \epsilon$.

An adversary gaining access to the device cannot retrieve the information required for device actuation, \textit{i}.\textit{e}., the computational task to be executed prior to its actuation. We have described a game-theoretic approach in Section~\ref{sec:analysis} that shows the overall probabilistic advantage of an adversary is negligibly small.

\subsection{Challenges and Solutions}
\label{subsec:Challenges and Solutions}
Implementing reliable ordering for device actuation is to provide a secure and self-executing state of devices at a pre-defined time for protecting the owner's privacy is deceptively non-trivial. Below, we discuss the challenges we encountered and describe how we addressed them.

\medskip\noindent\textbf{C1. Anonymous trigger from the hub to devices.} As mentioned before, we need to mask the channel activities, device activities, and the coupling between both the channel as well as the devices. Note that in this context, the encryption techniques merely hide the meaning of the message across the channel, not the fact that to which device this message belongs to, and hence, it reveals the user activity. In addition, the solutions-based on {\em traffic shaping}, which incur excessive communication and latency overhead, also fail to decouple the device to channel activities. {Note that in a traffic shaping solution, whenever a device wishes to communicate with other devices, it needs to send a fixed-size packet. Thus, the device activity leads to a specific communication pattern, by which the adversary can determine the user's activities, as well as, can learn which device has generated the message and which device has received the message.}

\fbox{\begin{minipage}{3.3em}
		\textit{Solution.}
\end{minipage}}
To address this challenge, the distribution of user-commands to each device in the home network is based on a pre-defined topology, \textit{e}.\textit{g}., ring, where a token rotates continuously within the one-directional (1-D) ring topology.\footnote{{\scriptsize In order to leverage a continuous channel-activity as a means to hide the actual channel-activity, the ring topology is efficient as compared to star alignment. In addition, one can also use the mesh-topology with anonymous-routing that requires asymmetric-key cryptography overheads at each relay-node while sending a token between a source and a destination.}} Thereby, channel activity remains consistent and independent of the devices actuated as a result of the workflow. {Thus, the adversary cannot learn which device has generated the message and which device has received the message. }In our context, the token ($\mathcal{T}$) has three fields: (\textit{i}) command field, which contains the encrypted commands corresponding to each device, (\textit{ii}) data field, which contains the device-generated data, and, (\textit{iii}) a toggle bit string field, which contains a $N$ bit string, where each bit denotes a unique device in the topology. This toggle bit string is used to indicate that a device is interested in uploading the freshly generated data during upcoming token arrival at the device. Since the order of device actuation reveals crucial information about the user activity inside the home, our token-based solution guarantees a secure ordering among devices while executing the commands. In addition, the device actuation is controlled in a manner that a recipient device itself cannot pre-decode and/or pre-pone the command execution.

\medskip\noindent\textbf{C2. Command execution and verifiable ordering.} After masking the channel activity through a constantly rotating token, the next step is to have a verifiable ordering of command execution at each device. Each device receives a command through the token. Now, these devices can decrypt and execute the command immediately. However, it again enables the channel to device activity mapping. Therefore, the next challenge is to insert an artificial delay between a device receiving the commands, and then, executing the commands at an appropriate time, without relating to any specific channel activity. The artificial delay enables a correct execution order at each comparable or non-comparable device.\footnote{{\scriptsize The comparable-devices are those that are defined under certain relative order in a workflow. The incomparable-devices are independent and are not restricted under any relative order with respect to other devices. }}

\fbox{\begin{minipage}{3.3em}
		\textit{Solution.}
\end{minipage}} The protocol message from the homeowner to hub includes the ordered commands ($c_{l}$) that pass through a device $D^i$ to another device $D^{i+1}$ using the anonymous token circulation. The token is encrypted using a shared symmetric key between the hub and the devices in the topology. %Further, the command field of the token occupies the partially ordered commands of length $l$, denoted by $c_l$.
The recipient device $D^i$ retrieves the encrypted command ($m^i$) from the command field of the token and begins with a puzzle computation. Note that the devices do not execute the commands immediately after receiving (time-locked) commands. In particular, these devices wait for a pre-defined amount of time before executing the command, such that neither the artificial delay at each device can be known by the adversary in advance, nor the devices can control this artificial delay to postpone or prepone the scheduled commands. However, this waiting period is not idle, and the devices resume on some computational task.

\medskip\noindent\textbf{C3. Anonymous response from devices to the hub.} Once the devices have received the commands in an anonymous manner, they may generate some data as a result of command execution. However, uploading this data immediately would reveal the device activity patterns. Clearly, this periodic channel activity (in the form of traffic) relates to a specific device that had executed the command recently. Therefore, the upstream data upload on the hub should be anonymous too.

\fbox{\begin{minipage}{3.3em}
		\textit{Solution.}
\end{minipage}} In our scheme, each device is required to send a request for data upload to the hub using the rotating token, which contains a toggle bit string field of size $N$ bits, where each bit represents a unique device in the topology. The $i^{\mathit{th}}$ toggled bit indicates that the device $D^i$ has generated the data and is ready to anonymously transmit the data by using the data field of the token. The data field is used to carry the device-generated data without revealing the data and the sender of the data. The data field contains random data as long as there is no request from the devices to send the data to the hub. The details of data upload phase from device to hub will be clear in step 4, given in Section~\ref{sec:decoup}.

\B
\subsection{Building Blocks}
This section provides a brief overview of basic building blocks used to develop the proposed solution as detailed in Section~\ref{sec:decoup}.

\medskip
\noindent\textbf{RSA puzzles.} The verifiable delay regarding device actuation is based on cryptographic RSA puzzles~\cite{rlock}. These time-bounded puzzles are useful for the applications that require security against the hardware parallelization attacks (\textit{i}.\textit{e}., bypassing a security solution by running a mathematical problem on different hardware in parallel) through Application Specific Integrated Circuits (ASIC). Accordingly, the puzzle solution is based on inherently sequential operations such as {\em modular exponentiation}. Let us assume that $t^{\prime}$ be the time to release an encrypted message. Also, a device is capable of computing $S$ number of square operations modulo $n$ per second. Thus, the puzzle requires sequenced exponentiations of ($a^{2^{\hat{t}}}$ {\em mod} $n$) where $n=pq$ is publicly known RSA modulus and $\hat{t}$, $p$, $q$ and $\phi(n)=(p-1)(q-1)$ remains secret. In particular, $\hat{t}=St^{\prime}$ denotes the difficulty level of the puzzle for a specific device. Therefore, the computation of this modular exponentiation operation requires either the inherently sequential execution of these operations {\em or} to solve the integer factorization problem.

%Breaking artificial delay is impossible.
In the proposed scheme, the RSA puzzles enable an artificial yet verifiable delay\footnote{{\scriptsize The delay for each device is computed by the hub, and done via device profiling, as explained in Section~\ref{sec:paray}.}} with respect to command execution. To compensate this delay an adversary must know the private key of a device and then invest the same time as the victim device was supposed to invest in, for the puzzle computation. In particular, the adversary can always lengthen the delay (which is easily detectable during the puzzle validity check), but cannot shrink the delay due to inherently sequential operations.

\begin{wrapfigure}{r}{0.53\textwidth}
\BBB\BBB\BBB%\BBB
	%\begin{framed}
	%\B%BB
	\begin{center}
		\includegraphics[scale=0.35]{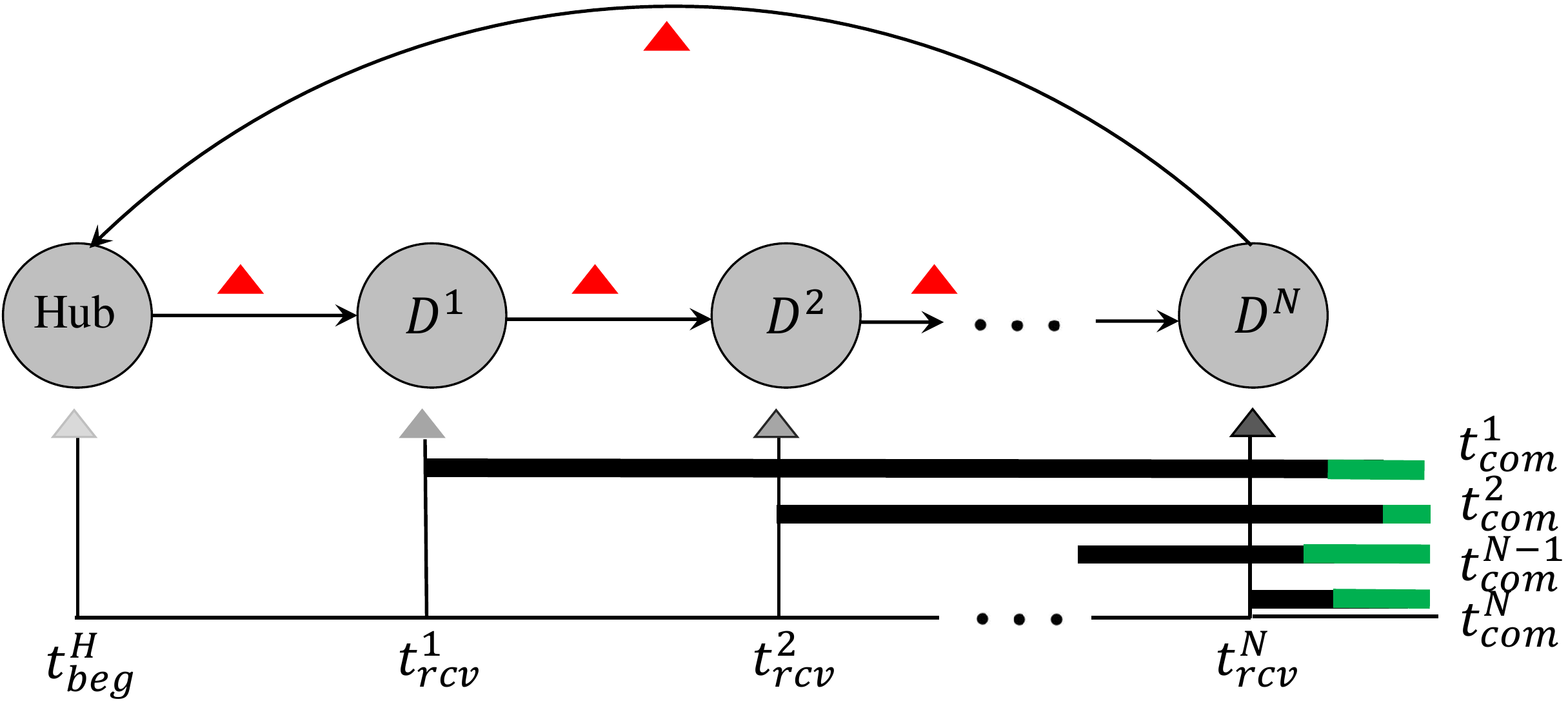}
	\end{center}
	%\vspace{0.2cm}
	%\end{framed}
	\BBB
	\caption{Device ordering on the timeline}
	%\BB%B
	\label{fig:problem statement}
	\B%BB\B%BB\B
\end{wrapfigure}

%\smallskip
\medskip
\noindent\textbf{Order-preserving bijection.} The order-preserving bijection guarantees an instance of the totally ordered set elements as derived from the partially ordered set of the same elements. In particular, the bijection provides a unique sequence of the totally ordered set elements. Let us assume there is a partial order relation $\mathcal{R}=\{{\leq}\}$ on set elements $\mathcal{E}=(e^1,e^2,e^3,e^4)$ such that $\mathcal{R}$ is: reflexive, \textit{i}.\textit{e}., $e^i\mathcal{R}e^i$; antisymmetric, \textit{i}.\textit{e}., if $e^i\mathcal{R}e^j$ and $e^j\mathcal{R}e^i$ then $e^i=e^j$; and transitive, \textit{i}.\textit{e}., if $e^i\mathcal{R}e^j$ and $e^j\mathcal{R}e^k$ then $e^i\mathcal{R}e^k$. In addition, the partially ordered set $\mathcal{E}^\prime=((e^1,e^2),(e^3,e^4))$ under relation $\mathcal{R}=\{\leq\}$ has a unique minimal and maximal element. Therefore, an order-preserving bijection generates a linear ordering of elements in set $\mathcal{E}^\prime$. Essentially, this linear extension generates a permutation order of the elements in a given partially ordered set $\mathcal{E}^\prime$. All of those permutation sequences in which $e^i$ appears before $e^j$ given that $(e^i\leq e^j) \: \in\mathcal{E}^\prime$ are a valid candidate as per any totally ordered set element sequence. The reliable ordering of device actuation is guaranteed through the linear extension of the owner-defined schedule even when the application host is unavailable.

\section{Decoupling Channel Activity from Device Activity}
\label{sec:decoup}
This section provides the details of our proposed protocol for decoupling channel activities from device activities. First, we illustrate the device-to-device interaction through an example as below:

\subsection{Example --- Device Ordering}
Consider $N$ number of devices ($D^1,D^2, \ldots,D^N$) that are connected through a hub ($H$), as shown in Figure~\ref{fig:problem statement}. Let $s_{\mathit{on}}^i$ (and $s_{\mathit{off}}^i$) be the on (and off) state of a device $i$. The owner $O$ can create a partial ordering for devices such as
$\langle(D^1,D^2,D^3),$ $(D^4,D^5),$ $\ldots,$ $(D^{N-1},D^N)\rangle$ based on their states, \textit{e}.\textit{g}., $\langle(s^1_{\mathit{on}},s^2_{\mathit{off}},s^3_{\mathit{off}}),$ $(s^4_{\mathit{off}},s^5_{\mathit{on}}),$ $\ldots,$ $(s^{N-1}_{\mathit{off}},s^N_{\mathit{on}})\rangle$
that shows the device $D^1$ must change its state to on, \textit{i}.\textit{e}., $s^1_{\mathit{on}}$, before the devices $D^2$ and $D^3$ change states to off, \textit{i}.\textit{e}., $s^2_{\mathit{off}}$ and $s^3_{\mathit{off}}$. Similarly, the device $D^4$ must change its state to off, \textit{i}.\textit{e}., $s^4_{\mathit{off}}$, before the device $D^5$ changes its state to on, \textit{i}.\textit{e}., $s^5_{\mathit{on}}$.\footnote{{\scriptsize For the sake of simplicity, this example includes the ordering between the devices and the corresponding states. However, throughout the paper, our focus is to order time intervals for devices' actuation.}} After creating the partial order of devices, the owner sends a message to the hub that sends the message (shown in red color) to one of the devices, as shown $D^1$ in Figure~\ref{fig:problem statement}. We refer to the message from the hub to devices as a token. Each device $i$ receives the token at time $t^i_{\mathit{rcv}}$, forwards the token, and begins computation at time $t^i_{\mathit{com}}$. The bottom part shows when each device receives the token in a sequence as ($D^1,D^2,\dots,D^{N-1},D^N$). Note that due to user-defined partial order of device actuation $\langle(D^1,D^2,D^3), (D^4,D^5),\ldots, (D^{\mathit{N-1}},D^N)\rangle$, the devices across the partial orders are mutually incomparable.\footnote{{\scriptsize If the user finds a change in his/her schedule, then another remote command can overwrite the previous commands and the schedule workflow, correspondingly.}} In Figure~\ref{fig:problem statement}, thick black line (for each device) shows that the device is having the token and waiting for the predefined time (given in the token) for its activation, and the green line shows when the devices start working.

\subsection{Verifiable Ordering Protocol}
This section formally defines the verifiable ordering protocol and a detailed description for each step as below.

\begin{definition}[Verifiable Ordering Protocol]
The verifiable device ordering protocol is a tuple $(\mathit{param},\mathit{puzgen},P_D^i,P_O)$ of four polynomial-time algorithms such that:
\end{definition}
\begin{itemize}[noitemsep,nolistsep,leftmargin=0.01cm]
\item \emph{Public parameter generator}. $\mathit{param}(1^{\lambda}) \rightarrow (n, a)$. $\mathit{param}(1^{\lambda})$ initializes the prime integer factors $n=pq$ (where $p$ and $q$ are two large prime numbers) and random value $a$ for the puzzle creation.

\item \emph{Puzzle generator}. $\mathit{puzgen}(n, a, \hat{t}, E_z, E_k) \rightarrow \mathcal{P}$. $\mathit{puzgen}(n, a, \hat{t}, E_z, E_k)$ selects the input values as target time for commands execution $\hat{t}$, encrypted command $E_z(z,k)$, encrypted key $E_k(a,\hat{t},n,k)$ and outputs a puzzle $\mathcal{P}$ for each device.

\item Follower ${D^i}(\mathcal{P}^i, SK) \rightarrow z(\hat{t}_i, t^i_{\mathit{com}}]$ completes the puzzle $\mathcal{P}^i$ and executes the command within the half-open interval, \textit{i}.\textit{e}., no earlier than $\hat{t}_i$ but earlier than or at $t^i_{\mathit{com}}$.

\item Owner $O(\mathcal{P}, \phi{(n)}, PK^i) \rightarrow \mathit{accept}(t^i_{\mathit{com}}\geq \hat{t}_i)$ accepts the timely command execution at each device using $\phi{(n)}$.

\end{itemize}

\noindent\textit{Setup and key distribution.} The manufacturing authority initializes a unique identity for the owner, hub, and, home devices by using the secure identity distribution function, say $\texttt{Init}(1^\lambda)\rightarrow \mathit{identity}$, where $1^\lambda$ is the security parameter that generates a unique $\mathit{identity}$ for each entity. The certificate authority verifies that each of these devices knows the private key paired to the public key proposed for certification as: the homeowner ($O_{\mathit{SK}},O_{\mathit{PK}}$), the hub ($H_{\mathit{SK}}, H_{\mathit{PK}}$), and $i^{\mathit{th}}$ home device ($D^i_{\mathit{SK}},D^i_{\mathit{PK}}$), possesses a valid key pair.

\LinesNotNumbered
\begin{algorithm}[t]
	\DontPrintSemicolon
	%\scriptsize
	%=========Inputs=========Inputs=========Inputs=========Inputs=========Inputs=========
	\textbf{Inputs:} set of $l\in N$ devices ($D^i$), public keys ($D^i_{\mathit{PK}}$)
	%$\mathcal{R}$: a relation having $n$ tuples and $m$ attributes, $c$: the number of non-communicating clouds
	
	%=================Variables=================Variables=================Variables=
	{\bf Variables:} $\mathcal{P}$ a puzzle, $z$ a command.
	%$t^i_{\mathit{rcv}}$ token receiving time, $t^i_{\mathit{fwd}}$ token forwarding time.
	%$\mathit{letter}$: represents a letter
	
	\nl{\bf Function $\mathit{create} \: (c_l,D^N)$} %\nllabel{ln:crtn}
	
	\nl \Begin{
		\nl \For{$\forall (i,j) \: \exists (D^i,D^j) \: \in H$}{
			%	\nl \lForEach{
			\nl $\mathit{schedule} \: ((D^i,D^{\mathit{i+1}}),(D^j,D^{\mathit{j+1}}))$
			
			\nl \For{$\forall (i,i+1)$}{      %for each subset order the commands
				
				\nl $z=(s_{\mathit{on}} \vee s_{\mathit{off}}) \wedge (\hat{t}_i \leq \hat{t}_{i+1})$
				
				\nl $\mathit{generate} \: (\mathcal{P}^i)=(n,a,\hat{t}_i,E_z,E_k)$
				
				\nl $m^i=enc(\mathcal{P}^i,D^i_{\mathit{PK}})$
				
				\textbf{end \: for}} %state on or off
			\textbf{end \: for}}
		\nl return $c_l=((m^i,m^{\mathit{i+1}}),(m^j,m^{\mathit{j+1}}))$
		\textbf{end}}
	\caption{Algorithm for order creation}
	\label{alg:crtn}
\end{algorithm}
\setlength{\textfloatsep}{0pt}

\medskip\noindent\textbf{Step 1: Order creation: schedule creation at the owner.} The homeowner $O$ first creates a schedule, say $\mathit{Schedule}$, for device actuation. The creation of schedule is inherently specific to the preferences of the owner on a day to day basis and can include all or a subset of the home devices. The schedule creation does not require interaction with any device $D^i$ or the hub $H$. Below we show a partially ordered timeline/schedule of four devices:

\centerline{$\mathit{Schedule}=((D^1,D^2),(D^3,D^4))$}

Where only the elements of the same subset are comparable, \textit{e}.\textit{g}., $D^1$ with $D^2$, and, $D^3$ with $D^4$, based on the timeline. The owner converts this schedule to a verifiable device ordering (see Algorithm~\ref{alg:crtn}), before sending it to hub. The function $\mathit{create}(c_l,D^N)$ of Algorithm~\ref{alg:crtn} converts a schedule for devices into the partially ordered sets of commands of length $c_l$, where $l\subseteq N$. Line 3 considers a pair\footnote{{\scriptsize We consider only pairs of devices, to simply demonstrate the relative ordering. However, a different subset may contain as large as the total number of devices.}} of devices in the topology of the hub. Line 4 creates a mutually dependent schedule for the pair of devices with temporal dependency. Lines 5 and 6 consider all of these mutually dependent pairs, decide the state of command as $z=(s_{\mathit{on}}\vee s_{\mathit{off}})$, and then, generate a unique puzzle $\mathcal{P}^i$ for each device in Line 7. Here, the puzzle message contains a tuple of variables $(n,a,\hat{t}_i,E_z,E_k)$, where $n$ is the product of two large prime numbers $p$ and $q$, $a$ is a random number, $\hat{t}_i$ is the time-complexity of the puzzle, $E_z$ is the encrypted command $z$ using key $k$, and $E_k$ is the encrypted key $k$. Line 8 encrypts each puzzle $\mathcal{P}^i$ into a message $m^i$ for a device $D^i$ using the public key of the device $D^i_{\mathit{PK}}$. Line 9 returns an assembled order $c_l$ that contains an encrypted message for each device corresponding to the mutually dependent devices in the schedule. This ends the creation of a relative order for the chosen set of devices. Next, the homeowner sends securely this order, say $\mathit{Order}$, to the hub, as follows:

\centerline{$\mathit{Order}=(O_{\mathit{id}},H_{\mathit{id}},c_l,\mathit{Sign}(\mathcal{H},O_{\mathit{SK}}))$}

\textit{i}.\textit{e}., the homeowner sends its identity ($O_{\mathit{id}}$), the identity of the hub ($H_{\mathit{id}}$), and encrypted order of commands ($c_l$) along with hash digest of all three attributes.

\begin{figure}[!t]
	%\begin{framed}
	\BBB%\BBB\BB
	\begin{center}
		\includegraphics[scale=0.4]{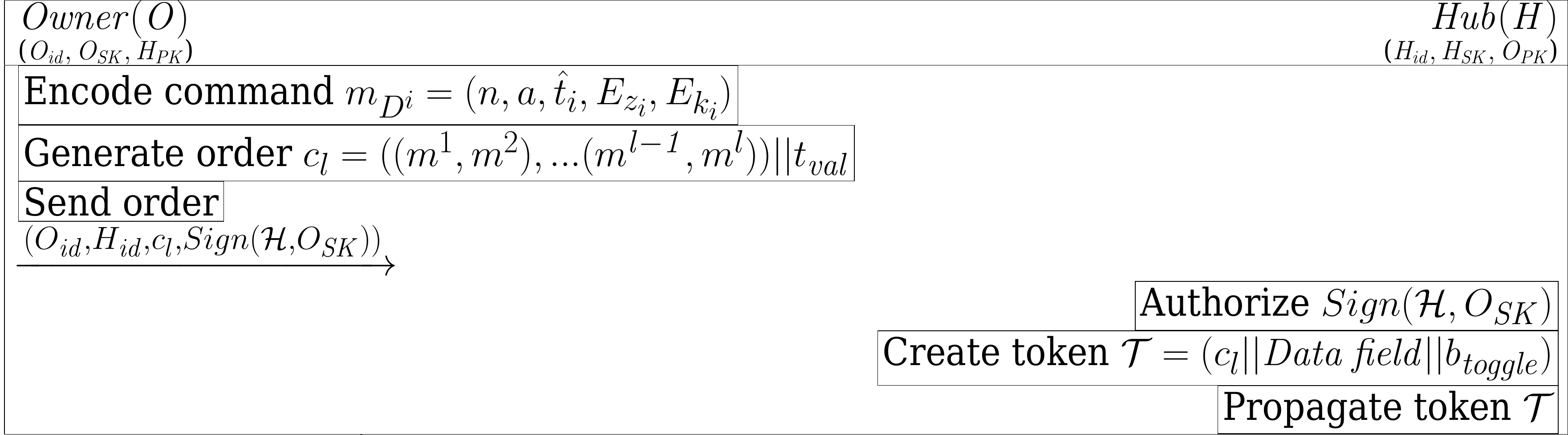}
	\end{center}
	%\vspace{0.2cm}
	%\end{framed}
\BBB
	\caption{Order creation}
	\label{fig:s1}
\end{figure}

\medskip\noindent
\textbf{Step 2: Token generation by the hub and token delivery to devices.} On receiving the partial order of commands from the owner, the hub verifies the sender by computing a local hash digest $\mathcal{H}^\prime$ over $(O_{\mathit{id}},H_{\mathit{id}},c_l)$. Also, the hub verifies the signature using $O_{\mathit{PK}}$ and compares the received hash digest $\mathcal{H}$ with the locally computed hash digest $\mathcal{H}^\prime$. If $\mathcal{H}=\mathcal{H}^\prime$ then the hub accepts this order.

After the verification of order origination, the hub creates a token that is used for order delivery (in this step) and for data collection generated by devices (step 4). The token has three fields: command field $c_l$, data field, and toggle bit string $b_{\mathit{toggle}}$. Every token field has sensitive information regarding the device activity. Therefore, we assume that the token is encrypted using a shared symmetric key $k_s$ among the devices and the controller hub: {$\mathcal{T}=E((c_l||\mathit{Data \: field}||b_\mathit{toggle}),k_s)$}

\begin{figure}[t]
\BBB\BB	%\begin{framed}
	\begin{center}
		\includegraphics[scale=0.4]{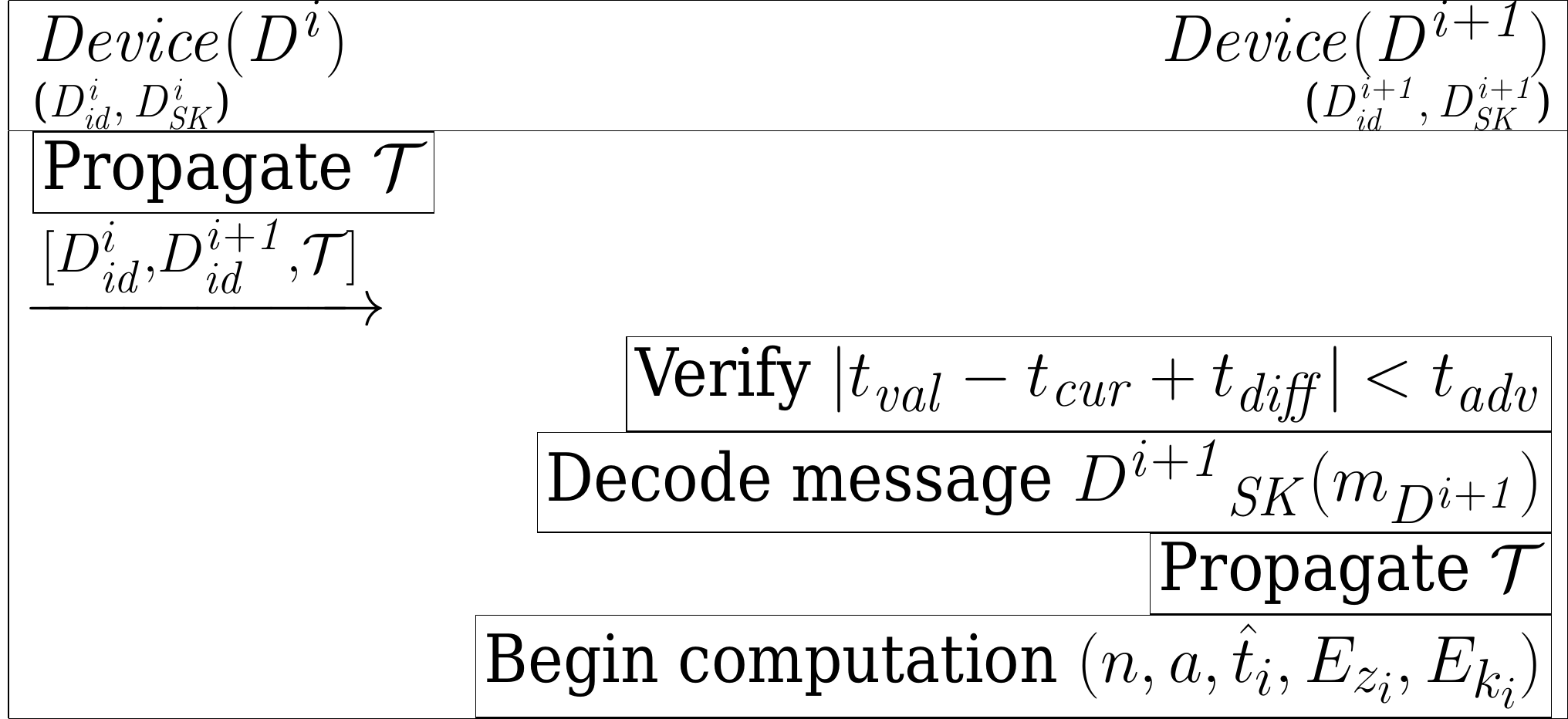}
	\end{center}
	%\vspace{0.2cm}
	%\end{framed}
\BBB
	\caption{Token circulation}
	\label{fig:s2}
\BBB\BB
	%\BBB\BBB
\end{figure}

The hub, then, forwards the token among all devices in the topology even if a device was not included in the schedule, as shown in Figure~\ref{fig:s1}. On receiving the token, each device decodes the corresponding command in token and forwards the token to the next peer device in the topology. The next peer device is chosen as per the topology underneath. Recall that the devices are connected in a pre-defined topology\footnote{{\scriptsize  It must be noted that from the practical deployment aspect it is difficult to connect these smart devices in a ring topology unless the devices belong to the same OEMs, \textit{e}.\textit{g}., Apple HomeKit. Therefore, a star or a grid topology can be used to combat the single point of failure and device heterogeneity in the current scenario.}} such as in a unidirectional ring, bidirectional ring, star (fault-tolerant), grid, mesh or hybrid setting. Note that the token rotates constantly in the topology (see Figure~\ref{fig:s2}).

\LinesNotNumbered
\begin{algorithm}[t]
	\DontPrintSemicolon
	%\scriptsize
	%=========Inputs=========Inputs=========Inputs=========Inputs=========Inputs=========
	\textbf{Inputs:} set of devices ($D^N$), user-defined schedule $((D^i,D^{i+1}),(D^j,D^{j+1}))$
	%$\mathcal{R}$: a relation having $n$ tuples and $m$ attributes, $c$: the number of non-communicating clouds
	%=================Variables=================Variables=================Variables=
	%{\bf Variables:} $t^i_{rcv}$ token receiving time, $t^i_{fwd}$ token forwarding time.
	%$\mathit{letter}$: represents a letter
	
	\nl{\bf Function $\mathit{chain}(D^i,D^j,\mathcal{R})$} \nllabel{ln:chng}
	\Begin{ 	
		\nl \For{$\forall (i,j,\mathcal{R}) \: \in \mathit{schedule}(D^i,D^j); \: i\mathcal{R}j=i<j$}{
			
			\nl $^nP_n(D^i)$
			
			\nl \For{$\forall \{D^i\}_{n!} \wedge i\mathcal{R}j=true$}{
				
				\nl return $^nP^{\prime}_n\{D^i\}^n_{\mathit{i=1}}$}
			
			\textbf{end \: for}}
		\textbf{end}}
	%  	
	%  	\nl \lForEach{$\mathit{generate} \: (\mathcal{P})=(n,a,\hat{t},E_z,E_k) \: z=s_{\mathit{on}}/s_{\mathit{off}}$}{$\hat{t}_u \leq \hat{t}_v \: If \: u \leq v$} %state on or off
	%  	}}
	\caption{Algorithm for chaining}
	\label{alg:chng}
\end{algorithm}
\setlength{\textfloatsep}{0pt}

Algorithm~\ref{alg:chng} explains the linear ordering of devices in function $\mathit{chain}(D^i,D^j,\mathcal{R})$. The linear ordering condition requires that each device $D^i$ and $D^j$ must satisfy: the exact same mutual ordering or relation $\mathcal{R}=\{\leq\}$ as in $c_l$. Line 2 selects each pair ($i,j$) of device that is paired under relation $\mathcal{R}=\{\leq\}$ in the original schedule. Line 3 enumerates all possible permutations of these devices, say $^nP_n(D^i)$ where $i\in N$. Line 4 selects one permuted order $^nP^{\prime}_n$ of devices (from the total number of permutations) such that the precedence relation still holds true, \textit{i}.\textit{e}., $i\mathcal{R}j=true$. Finally, in line 5, the selected topological order $^nP^{\prime}_n$ is returned from all possible permuted orders $^nP_n(D^i)$ of the devices.

%\vspace{0.2cm}

\begin{figure}
	\begin{center}
		\fbox{
			\begin{minipage}[t]{6.5in}
				%\footnotesize
				\noindent
				\normalfont
				\textbf{The puzzle messages}:
				%\vspace{-0.2cm}
				\begin{center}
					\begin{tabular}{l l l}
						$n=pq$ & & \\
						
						$\hat{t}=St^{\prime}$ & & \\
						
						$E_z=\mathit{enc}(z,k)$   & &    \\
						
						$E_k=k+{a^{2^{\hat{t}}}} \ (mod \ n)$     &    &  \\
					\end{tabular}
				\end{center}
				\noindent
				\textbf{The puzzle computation ($\mathcal{P}$)}:
				%\vspace{-0.2cm}
				\begin{enumerate}[noitemsep,nolistsep]%,leftmargin=0.2in]
\item Initially, $\mathcal{P}$ receives as input a secret key $k$ and encrypt the original message $z$ denoted as $E_z$. It must be noticed that each individual command corresponding to a device $D^i$ is secured in the form of an encrypted message $m_i$. Furthermore, a cascaded command as a whole contains multiple messages of these types.
					%Also the puzzle computation is an injective function such that each individual message intended for a specific device.
					
\item Subsequently, $\mathcal{P}$ receives inputs as a secret key $k$, random number $a$, puzzle difficulty level $\hat{t}$, modulus $n$, and then generate $E_k$. The puzzle computation relies on the secrecy of key $k$ used to encrypt a secret message $E_z=(z,k)$. Also, $\hat{t}=S*t^{\prime}$ denotes the difficulty level of the puzzle for a specific device, which can perform $S$ number of square operations per second and $t^{\prime}$ is the time to decrypt the message using a regular encryption scheme.
					
\item The puzzle $\mathcal{P}$ includes the tuple ($n,a,\hat{t},E_z,E_k$) for which the computation task is to be solved. Evidently, the recipient of this puzzle would have to spend at least $\hat{t}$ amount of time to complete the computation task and reveal the key $k$.
				\end{enumerate}
			\end{minipage}
	}\end{center}
\BBB
	\caption{Puzzle computation}
	\label{fig:s5}
\end{figure}

\medskip
\noindent\textbf{Step 3: Order retrieval and puzzle computation at the devices.} Note that the activation sequence of devices is released ahead of their actual activation; however, we need to restrict them not to execute the command before the prescribed time. Thus, in this step, devices perform a pre-defined computation task (as detailed in Figure~\ref{fig:s5}) to unlock and execute the owner-defined command.

It must be noted that unlocking the command is as necessary as unlocking the command within the prescribed duration, \textit{i}.\textit{e}., knowing the particular order of a device in the overall sequence. A device receives the computation task in the form of a puzzle as soon as the order is delivered in step 2. Subsequently, the device begins the computation task if the puzzle validity period has not expired yet. The time-bound during which a device is restricted to begin, as well as, end the computation task cannot be compressed unless the device possesses a distinguisher for the factoring problem. Therefore, the verification that the secure computation task is crucial. This can be verified through the Euler's totient as a trapdoor for factoring $n$ inside the puzzle. Note that the value of $n$ is public and the value of $\phi{(n)}$ is kept secret. Therefore, computing $\phi{(n)}$ from $n$ is as hard as integer factorization. In addition, without the knowledge of $\phi{(n)}$ the computation time for $E_k$ is directly dependent on $\hat{t}$ time-consuming square operations.

%Each device $D^i$ retrieves the corresponding puzzle $\mathcal{P}^i$ and start computing the puzzle.

\LinesNotNumbered
\begin{algorithm}[t]
	\DontPrintSemicolon
	%\scriptsize
	%=========Inputs=========Inputs=========Inputs=========Inputs=========Inputs=========
	\textbf{Inputs:} a puzzle ($\mathcal{P}$) with a set of public variables ($n,a,\hat{t}_i,z_i,k_i$)
	%$\mathcal{R}$: a relation having $n$ tuples and $m$ attributes, $c$: the number of non-communicating clouds
	
	%=================Variables=================Variables=================Variables=
	%{\bf Variables:} $t^i_{\mathit{rcv}}$ time at which token was received, $t^i_{\mathit{fwd}}$ time at which token was forwarded.
	%$\mathit{letter}$: represents a letter
	
	\nl{\bf Function $\mathit{verify}(D^i,\hat{t}_i)$} \nllabel{ln:vrfy}
	
	\nl \Begin{
		\nl  	 \For{$\forall (i,j) \in\mathit{schedule}((D^i,D^{i+1}),(D^j,D^{j+1}))$}{
			\nl \If{${a^{2^{\hat{t}}}} \: mod \: n \equiv {a^{2^{\hat{t}} \: mod \: \phi{(n)}}} \: mod \: n$}{
				\nl \If{$(t^i_{\mathit{com}} \le t^{i+1}_{\mathit{com}}) \wedge (t^j_{\mathit{com}} \le t^{j+1}_{\mathit{com}})$}{
					
					\nl	return $\mathit{True}$}
				\textbf{end \: if}}
			\textbf{end \: for}
		}return $\mathit{False}$ \textbf{end}}
	%\nl \For{$\forall \: i \: \exists D^i \: \in H$}{
	%	\nl \lForEach{$\mathit{compare}(t_{\mathit{rcv}}^i,t_{\mathit{fwd}}^i,\hat{t})$}{
	%\nl \lIf{$\hat{t} \geq (t_{\mathit{fwd}}^i - t_{\mathit{rcv}}^i)$}{
	%	\nl \For{$\forall (i,j) \in \mathit{schedule}((D^u,D^v),(D^x,D^y))$}{$\mathit{chain}(i,j)$}}}%order-preserving bijection
	%}}
	\caption{Algorithm for delay verification}
	\label{alg:vrfy}
\end{algorithm}
\setlength{\textfloatsep}{0pt}

Algorithm~\ref{alg:vrfy} explains the verification of time-bounded commands in function $\mathit{verify}(D^i,\hat{t}_i)$. Line 3, considers all devices that are part of the current schedule. In order to verify that a specific device has executed the command within the pre-defined interval, $O$ securely pre-computes the Euler's totient $\phi{(n)}=(\mathit{p-1})(\mathit{q-1})$ such that $p,q$ is discarded after computing the $n$ and $\phi{(n)}$. Line 4, verifies the time bound for each of these devices, such that

\centerline{${a^{2^{\hat{t}}}} \: mod \: n \equiv {a^{2^{\hat{t}} \: mod \: \phi{(n)}}} \: mod \: n$}

Line 5 compares the execution order of devices that have happened as a result of puzzle computation and returns true in line 6 if it is a total order.

\medskip\noindent
\textbf{Step 4: Data generation at the devices.} Once the devices have completed the puzzle computation, they generate the data as a result of the command execution. These home devices are bound to upload the locally generated data to the hub. In our scheme, the token contains an anonymous data field to securely transmit the device-generated data to the hub. We use bitwise ($b$) XOR padding to overwrite the random data in the token data field as:

\centerline{$\mathit{Overwrite} \: \mathit{data} \: (b_o)=\mathit{Random} \: \mathit{data} \: (b_r) \xor \mathit{Generated} \: \mathit{data} \: (b_g)$}

It is cryptographically hard to distinguish the presence of random data from the device-generated data as stored inside the token. Note that our threat model does not consider the ISP or DNS level threats, therefore, devices are only assumed to securely generate and anonymously dispatch the data to the hub and combat any passive learning attacks within the physical periphery of the home.

\noindent\textit{Collision:} Our token circulation strategy and the token structure are primarily for smart home scenarios, where we assume that the single token field can accommodate the peak hour traffic. However, when the peak hour traffic exceeds and multiple devices request for data upload (for example, in a multi-tenant building), more data fields are required to avoid the collision situation. A simple approach is to create sub-fields inside the data field such that each sub-field belongs to a unique device. Thus, each device can fairly utilize the data upload capacity in any round during the token circulation.

\subsection{Time Analysis}
The time spent during the token circulation and puzzle computation is directly proportional to the number of devices connected in the network. For example, token begins at time ($t^H_{\mathit{beg}}$) at the hub and completes the first round of token circulation at time ($t^H_{\mathit{end}}$). The time spent at $i$th device is ($t^i_{\mathit{fwd}}-t^i_{\mathit{rcv}}$) that receive the token at time ($t^i_{\mathit{rcv}}$) and forward it to next device at time ($t^i_{\mathit{fwd}}$). Therefore, the total time spent in one round of token circulation:
$$t_{\mathit{sum}}=(t^H_{\mathit{end}}-t^H_{\mathit{beg}})-(t^i_{\mathit{fwd}}-t^i_{\mathit{rcv}})_{\mathit{i=1}}^N$$
Note that the token circulation time is sequenced and linear w.r.t. the number of devices. While the puzzle computation time $t^i_{com}$ varies independently among all devices. So the puzzle computation time at $i$th device is $t^i_{\mathit{com}}\approx \hat{t}^i$. In order to optimize the puzzle computation time and still retain the verifiable guarantees on the artificial delay, we consider two types of device timelines.
  \begin{itemize}[noitemsep,nolistsep,leftmargin=0.01cm]
  	\item \textit{For comparable devices:} Each pair of comparable devices in the topology requires that $\hat{t}$'s are {\em at least} $(N-1)(t^N_{\mathit{fwd}}-t^{N-1}_{\mathit{fwd}})$ apart. The devices forward the token before beginning the local computation task. Any two adjacent devices ($D^{N-1},D^N$) that begin the computation after forwarding the token, must possess:
  $$|\hat{t}^{N}-\hat{t}^{N-1}|\geq(N-1)(t^N_{\mathit{fwd}}-t^{N-1}_{\mathit{fwd}})$$
    	\item \textit{For incomparable devices:} The set of incomparable devices require that $\hat{t}$'s are {\em exactly} $(j-i)(t^j_{\mathit{fwd}}-t^{i}_{\mathit{fwd}})$ apart. Every time a device $D^i$ forwards a token to $D^{i+1}$ it jumps $(t^{i+1}_{\mathit{fwd}}-t^{i}_{\mathit{fwd}})$ ahead on the computation timeline with respect to next device due to token propagation delay. Therefore, in order to provide an identical time of actuation for all incomparable devices:
  $$|\hat{t}^{i}-\hat{t}^{j}|=(j-i)(t^j_{\mathit{fwd}}-t^{i}_{\mathit{fwd}})$$
  	%\item \textit{For both comparable and incomparable devices:}
  	\end{itemize}

The total number of slots required is $(N-k)+1$ where $k$ represents the number of comparable devices. In particular, each comparable device requires a unique and non-overlapping $|\hat{t}|$ w.r.t. other comparable devices; while each incomparable device can be scheduled for an identical and overlapping $|\hat{t}|$.

\noindent\textit{Token frequency:} The token frequency is a crucial attribute from the perspective of how early a user can decide the schedule for all $N$ devices and, how many data upload requests are received during the peak hours. The frequency of token circulation can be either {\em fixed} or {\em random}. Let us assume a {\em fixed} slot $i$ between any two consecutive rounds of the token circulation such that the token begins a new round at every $i$th unit of time. The optimal length of the slot is the same as the maximum $\hat{t}^i$ in any schedule.
$$\mathit{Slot \: length}=max\{\hat{t}^i\}_{i=1}^{l\in N}$$
For example, if $(\hat{t}^1,\hat{t}^2,\dots,\hat{t}^{l-1},\hat{t}^l)$ is the time-bound for $l$ devices in any scheduled workflow then the slot length is same as the farthest possible device on the timeline of a scheduled workflow. %If $T$ is the length of the timeline for which the slot length is to be optimized then there will be as many number of slots as $T \: mod \: max\{\hat{t}^i\}_{i=1}^{l\in N}$.

   \begin{table*}[h]
   	\BBB\B
 	\begin{center}
 	%\begin{spacing}{0.35}
 	\begin{tabular}{l l l l}
 	\hline
 	%            \multicolumn{10}{|c|}{Country List} \\
 	%            \hline
 	{Protocol} & {Cost at device} & {Cost at hub} & {Cost at owner}  \\ \hline
 	Proposed scheme & ($2D+1XOR+\hat{t}Me$) & ($1S+1E+1XOR$) & ($1S+lE+\hat{t}mod{\phi{(n)}}Me$) \\
 	Scheme~\cite{tifs} & ($3H+7XOR+1E+1D$) & ($5H+8XOR+1E+1D$) & - \\
 	Scheme~\cite{shen} & ($3M+2H+4XOR$) & ($1H+4XOR$) & - \\
 	Scheme~\cite{thir} & - & ($Cp_m+Tp_m^{\prime}$) & - \\ \hline
 	%Scheme~\cite{} & ($9H+1XOR+1D$) & ($9H+3XOR+2E$) & - \\ \hline
 	\end{tabular}
 %	\F
 	\caption{Cost comparison between our scheme and existing schemes~\cite{tifs,shen,thir}}
 	\label{table:cost}
 	\BBB\BBB\BB\B
 	 	%\end{spacing}
 	\end{center}
 \end{table*}

Table~\ref{table:cost} represents the cost comparison based on mathematical operations such as encryption ($E$), decryption ($D$), signature generation and verification ($S$), exclusive-OR ($XOR$), hashing ($H$), scalar multiplication ($M$), and modular exponentiation ($Me$). The scheme in~\cite{thir} imposes a relative overhead such that ($p_m$) number of masking packets are required per traffic flow in case the traffic flow is lesser than a pre-defined threshold value. Therefore, the total overhead per traffic flow is ($Cp_m$) where $C$ is the communication overhead per packet. Similarly, if the traffic flow is above the threshold value then those excess packets $p_m^{\prime}$ are delayed and stored inside a queue. Therefore, the total latency per traffic flow is ($Tp_m^{\prime}$) where $T$ is the latency overhead per packet. As shown here that our proposed scheme requires the minimum number of operations. Further, in our approach the computational complexity at devices is variable and it depends on the required number of modular exponentiations, \textit{e}.\textit{g}., $\hat{t}$, as initialized by the owner.

\section{Security Analysis}\label{sec:analysis}
This section provides the security analysis for the proposed scheme. We first model the security experiment, below, like the standard security model.

  \begin{attackgame} Let $\mathcal{I}$ be the order-preserving protocol between the challenger and adversary $\mathcal{A}$ then the attack game works as:
  \end{attackgame}
  \begin{itemize}[noitemsep,nolistsep,leftmargin=0.01cm]
  	\item Public parameter generation: The challenger generates $(n,a)$ using $\mathit{param}(1^{\lambda})$.
  	\item Puzzle generation: The challenger generates $\mathcal{P}$ using $\mathit{puzgen}(n, a,$ $\hat{t}, E_k, E_z)$.
  	\item Query phase: An adversary attempts to attack $\mathcal{I}$ through token query given the access to a recently generated token $\mathcal{T}^{\prime}$. The adversary sends a value $\hat{t}$ to the challenger. The challenger generates the corresponding puzzle $\mathcal{P}$ and adds in $c_l$. The follower devices receive $c_l$, extract the unique puzzle, and execute the command. The challenger then sends $\mathcal{T}^{\prime}$ to the adversary.
  	\item State identification attempt: The adversary attempts to retrieve the intermediate state of the computation task and attempts to solve the puzzle earlier than the device through $t^{\mathcal{A}}_{\mathit{com}}$ for the same $\mathcal{P}$ in $\mathcal{T}^{\prime}$, such that $t^{\mathcal{A}}_{\mathit{com}}$ is lower than the original $\hat{t}$.
  \end{itemize}
An adversary $\mathcal{A}$ wins the game, if $t^{\mathcal{A}}_{\mathit{com}}<\hat{t}_i$ and the owner outputs $\mathit{accept}$. The probabilistic advantage of the adversary, $\mathit{Adv(\mathcal{A})}$, for winning the game is:
$$\mathit{Adv(\mathcal{A})}=Pr[t^{\mathcal{A}}_{\mathit{com}}<\hat{t}_i]$$
%For instance, the probability of guessing the total-ordering is as uncertain as the number of incomparable-devices considered in the permuted-orders:
%$$Pr\left( \frac{\mathit{selected \: order}}{\mathit{all \: permuted \: orders}}\right)$$
%this would give a contaminated view as proportional to the number of incomparable-devices considered in the permuted-orders.

We present a sequence of games as $\texttt{Game 0}$ to $\texttt{Game 2}$. Each $\texttt{Game i}$ shows that the advantage of an adversary $Pr[t^{\mathcal{A}}_{\mathit{com}}<\hat{t}_i]$ is negligibly small. Similarly, each subsequent game $\texttt{Game (i+1)}$ is produced through the previous game, such that the changes in secret parameters remain indistinguishable to the adversary. Therefore, the advantage of an adversary through changes in secret parameters (\textit{i}.\textit{e}., transition ing from one game to another) remains negligibly small. If $Pr[\mathcal{A}(i) \rightarrow 0]-Pr[\mathcal{A}(i+1) \rightarrow 0]$ is non-negligible then that adversary can be used as a distinguisher or as a solver for the integer factorization problem in our scheme; where $Pr[\mathcal{A}(i)]$ and $Pr[\mathcal{A}(i+1)]$ represent the probability of adversary winning the $\texttt{Game i}$ and $\texttt{Game (i+1)}$, respectively. The $\texttt{Game 0}$ represents the original attack such that $(t^{\mathcal{A}}_{\mathit{com}}=\hat{t}_i)$ and the artificial delay before the command execution is kept null. The $\texttt{Game 1}$ represents the attack with enhanced $t^{\prime}$ while $(t^{\mathcal{A}}_{\mathit{com}}<t^\prime\wedge\hat{t}_i)$. Similarly, the $\texttt{Game 2}$ represents the attack with general $t^{\prime}$ while $(t^{\mathcal{A}}_{\mathit{com}}=t^\prime)$ but $(t^{\mathcal{A}}_{\mathit{com}}<\hat{t}_i)$.

  \noindent\texttt{Game 0:} [\textbf{Record attack}] Let us assume that the puzzle $P_1$ contains $\hat{t}=0$ then the device must perform only one iteration to compute and decode the enciphered command. However, an adversary cannot distinguish an early puzzle such as $P_1$ from a delayed puzzle such as $P_2$ for which $\hat{t}>0$. In the token query phase, an adversary gathers the token transcripts for a known value of $\hat{t}$.

  {\upshape
  	\small
  	\ttfamily
  	\begin{tabbing}
  	Exp\=eri\=ment $\aexp^{t^{\mathcal{A}}_{\mathit{com}}=\hat{t}_i}$\\
  	\> let $c_l((m^1,m^2),\ldots,(m^{l-1},m^l)) \leftarrow \aaa(\mathcal{T}) $\\
  	\> generate $m^i(\hat{t})$ at random\\
  	\> $(m^k(\hat{t})) \leftarrow \aaa(c_l((m^1,m^2),\ldots,(m^{l-1},m^l)))$\\
  	\>if ($m^k(\hat{t})=m^i(\hat{t})$) \\
  	\> \> return $1$\\
  	\> else return $0$
  	\end{tabbing}
  }

It is computationally hard to distinguish between encrypted commands and to identify the command that carries known $\hat{t}$ within time $t^{\mathcal{A}}_{\mathit{com}}=\hat{t}_i$. An adversary can distinguish the commands with the advantage
$$\mathit{Adv(\mathcal{A})}_{\texttt{Game 0}}=Pr[m^i(\hat{t})\leftarrow c_l((m^1,m^2),\ldots,(m^{l-1},m^l))]\leq \epsilon$$

  \noindent\texttt{Game 1:} [\textbf{Clone attack with lesser $t^{\prime}$}] Let us assume that the adversary computes the puzzle in time $t^{\mathcal{A}}_{\mathit{com}}<t^\prime \wedge \hat{t}_i$ where the unit time capacity $t^{\prime}$ of the adversary is slower. Therefore, the advantage of the adversary depends on the probability to compute the $\hat{t}$ square operations faster than the home device. This requires that the adversary can solve the prime factors for modulus $n$.
  $$\mathit{Adv(\mathcal{A})}_{\texttt{Game 1}}=Pr[(p,q)\leftarrow(n)]\leq \epsilon$$   % \\ impossible in ideal world

  \noindent\texttt{Game 2:} [\textbf{Clone attack with general $t^{\prime}$}] Let us assume that the adversary can compute as fast as the home device, \textit{i}.\textit{e}., $t^{\prime}$. In particular, the adversary can also perform $S$ number of square operations per second. The probability $Pr[t^\prime=t^{\mathcal{A}}_{\mathit{com}}<\hat{t}_i]$ that an adversary $\mathcal{A}$ can solve a puzzle with the difficulty level $\hat{t}$ in lesser time than the home device is negligibly small. Since the adversary must compute $S^{\prime}$ number of operations for each of $S$ operations at home device, where ($S^{\prime}-S>\epsilon$):
     $$\mathit{Adv(\mathcal{A})}_{\texttt{Game 2}}=Pr[t^{\mathcal{A}}_{\mathit{com}}(S^{\prime}t^{\prime})\leftarrow \hat{t}_i(St^{\prime})]\leq \epsilon$$     %  \\  conditional attack on h/w or computational capacity

In this sequence of games $\texttt{Game 0}$ through $\texttt{Game 2}$ the total advantage of the adversary $\mathit{Adv(\mathcal{A})}$ depends on the sum of the probability to win each of these games.
$$Pr[t^{\mathcal{A}}_{\mathit{com}}=\hat{t}_i]+Pr[t^{\mathcal{A}}_{\mathit{com}}<t^\prime \wedge \hat{t}_i]+Pr[t^\prime=t^{\mathcal{A}}_{\mathit{com}}<\hat{t}_i]\leq \epsilon$$
Overall, the advantage of the adversary is proportional to the availability of computational resources to solve the puzzle for all devices in parallel. Similarly, the advantage of the adversary with respect to a single puzzle and a single device is proportional to the availability of computational resources to solve the inherently sequential operations of that individual puzzle. Therefore, the total advantage of an adversary to clone the entire timeline depends on the total computational power for both, the parallel and the sequential operations to decrypt the command ahead of time.

\begin{table}[!t]
\BBB\BBB\BBB
	\begin{center}
	%\begin{spacing}{0.35}
	\begin{tabular}{l l l l l l}
	\hline
	%            \multicolumn{10}{|c|}{Country List} \\
	%            \hline
	{Properties} & \cite{tifs} & \cite{shen} & \cite{thir} & Our scheme  \\ \hline
	Upstream direction & $\checkmark$ & $\checkmark$ & $\checkmark$ & $\checkmark$ \\
	Downstream direction & $\checkmark$ & $\times$ & $\checkmark$ & $\checkmark$ \\
	Verifiable delay & $\times$ & $\times$ & $\times$ & $\checkmark$ \\
	Partial ordering & $\times$ & $\times$ & $\times$ & $\checkmark$ \\
	Total ordering & $\times$ & $\times$ & $\times$ & $\checkmark$ \\
	Privacy & $\times$ & $\times$ & $\checkmark$ & $\checkmark$  \\
	Passive attack resistant & $\times$ & $\times$ & $\checkmark$ & $\checkmark$ \\
	Active attack resistant & $\checkmark$ & $\checkmark$ & $\times$ & $\checkmark$ \\ \hline
	\end{tabular}
	%\F
	\caption{Comparison between our and existing schemes}
	\label{table:cmp}
\BBB\BB
	%\end{spacing}
	\end{center}
\end{table}

Table~\ref{table:cmp} shows a comparison between our proposed scheme and the existing work. The comparison is based on the data flow direction, ordering, verification of ordering, privacy violation, and attack resistance.

\section{Extensions}
\label{sec:extend}
Our proposed solution (as mentioned in Section~\ref{sec:decoup}) that avoids any inference attacks regarding the device workflow is based on a static ring topology. However, the in-home device communication scenarios can be perceived as a trade-off between efficiency and privacy requirements. The solution detailed in Section~\ref{sec:decoup} can be used for applications that require stronger privacy guarantees at the cost of communication resource overhead. The ring topology has limitations, as follows:
\begin{itemize}[noitemsep]
  \item As the number of devices in the ring topology increases, the token size and the communication \emph{latency} also increase. By the latency, we refer to the time taken by a token to complete a single round in the given ring network.
      As we will see in Section~\ref{sec:Experimental Evaluation} (Figure~\ref{fig:latency_exp}), the mean latency grows linearly as the ring size is 39 and beyond that mean latency grows faster and consumes up to 250 milliseconds on a ring of size 75.

  \item Different workflows include heterogeneous devices with different characteristics in terms of the frequency with which devices need to communicate. For example, a garage door opener typically needs to communicate a few times a day, while a motion sensor communicates more frequently. A ring topology requires all devices to communicate at the highest frequency required by any device, in order to preserve privacy guarantees.

  \item The ring topology is susceptible to failures, when a single device switches off or leaves the network.
\end{itemize}

In this section, we extend our basic strategy to overcome the above-stated limitations. In particular, Section~\ref{sec:paray} provides a method that can handle a large number of devices in the network, while not increasing the latency and the token size. The proposed solution, as will be given in Section~\ref{sec:paray}, achieves scalability at the cost of a limited leakage that does not reveal the user's privacy, as well as, the privacy of workflow. Section~\ref{sec:sly} provides ways to handle different-sized data demand by different devices. We have also adopted a flower topology wherein each device communicates via the hub. {As a result of the flower topology, will be presented in Section~\ref{sec:flow}, devices leaving a network do not cause other devices to be blocked. However, the departure of the devices may reveal the user privacy, and to prevent this, we provide a solution based on the transmission of phantom messages to ghost devices.}

%\subsection{Parallel rings through device pre-profiling}
\subsection{Scalability via Parallel Rings}
\label{sec:paray}
The primary concern is to stabilize the overheads in terms of the token size and communication latency, irrespective of the growing ring size. We propose to cluster devices into multiple groups based on the frequency of their communication in the workflows. Each cluster possesses a ring topology (under basic construction in Section~\ref{sec:decoup}) or a modified ring, \textit{i}.\textit{e}., flower topology (under failure resilient construction, as we will see in Section~\ref{sec:flow}). Further, the clusters share a common hub. By executing several rings in parallel, it reduces the token size and latency overheads.

{This extension, however, comes at the cost of limited disclosure. Recall that the token rotates continuously in the topology, as mentioned in Section~\ref{subsec:Preventing Inference Attacks}. However, as the number of devices increases, the continuous token circulation may increase the latency and consume significant bandwidth. Thus, we can partition the devices into multiple rings, and it reduces the latency of each ring. However, even when partitioning devices into multiple rings, the continuous token circulation in each ring may consume significant bandwidth. In order to save bandwidth, we can selectively rotate the token in each ring, depending on the need of commands and data upload requests by devices. Based on these communication patterns, the adversary may classify the devices into low- or high-frequency devices. The devices that participate in the token circulation more frequently, we call them high-frequency devices (in terms of communication patterns); and the devices participating in the token circulation not often, we call them low-frequency devices.}

The parallel rings can also be leveraged into an application-specific adaptive ordering on top of the user-defined flexible ordering of command execution. Recall that in the current approach, devices can receive the command at any time, and then, execute the desired command without regarding the resource availability. For example, the cloud connectivity may be considered as a resource~\cite{doan}, and these devices may schedule the commands only when the cloud connectivity is most likely available.
While commands must be executed as per the user-defined schedule, an additional logic for device ordering can achieve better utilization of resources in the parallel ring execution, compared to the single ring scenario (will be clear soon). The decision to select a parallel or a sequential ring execution can be simply derived from the resource availability that is a derivative of load variations during a certain interval. For example, $\delta$ number of devices might increase during peak load interval, [$t_i, t_j$], and decrease during non-peak load interval [$t_j, t_k$]. %In order to create parallel ring while also supporting an application-specific ordering of devices, below, we define the concept of device profiling.
In order to enable parallel rings while also optimizing the computation power and energy resources, below, we present an extension to the proposed scheme of Section~\ref{sec:decoup} based on  {\em device pre-profiling}.

\medskip
\noindent\textbf{Device profiling.} Device profiling is based on the assumption of a model that detects the availability of resources  and different cost factors. For example, \emph{computation cost} depends on the power required to complete the compute operation; \emph{bandwidth cost} depends on which channel is selected `or' what time a channel will be selected for transmission. We assume that this resource availability window is sustained for an enough amount of time; for example, daily or weekly basis, \textit{i}.\textit{e}., the cost variations are defined for a day or a week. Therefore, based on device profiling, we can create a different ordering among the devices to achieve an optimal cost for these resources, while retaining the user-defined ordering. However, this adaptive command execution can reduce the operational cost only if: ({\em i}) the resource availability model is known to the hub, ({\em ii}) the user-defined schedule is available at the hub, and ({\em iii}) a logic is available at the hub that takes device profiles as input and generates a cost-efficient ordering of command execution.

Thus, the key idea is to create a profile $\varmathbb{P}$ for each registered home device that connects with the in-home network. The device profile $\varmathbb{P}_{D^i}=(\rho_{D^i}, \gamma_{D^i}, \Re_{D^i})$ for a device $D^i$ includes the following three values:
(\textit{i}) $\rho_{D^i}$: is the number of commands $m^i_j$ received per day at the device $D^i$ as $\rho_{D^i}=(m^i_1, m^i_2,\ldots, m^i_j)$, where each command includes the time when it needs to be executed at $D^i$.
(\textit{ii}) $\gamma_{D^i}$: is the dependency of $D^i$ on a set of $j$ devices (\textit{i}.\textit{e}., $\gamma_{D^i}=(D^1, D^2,\ldots, D^j)$), such that all commands $m^j$ at $D^j$ must be executed before the command execution at $D^i$.
(\textit{iii}) $\Re_{D^i}$: is the resources available at the device $D^j$, and based on $\Re_{D^i}$, the hub tunes the artificial delay for the device $D^i$.\footnote{{\scriptsize
{Device profiles can be created either by the user or by using some tools developed by industries, such as SystronicsRF ({\url{https://www.systronicsrf.com/technology/device-profiling.html}}). However, it needs integration of such tools with the hub, thereby the hub can use these profiles to establish multiple rings. However, the creation of the device profile is outside the scope of this paper.}}}

\begin{figure}[t]
\BBB\BBB\BBB
	%\begin{framed}
	\begin{center}
		\includegraphics[scale=0.4]{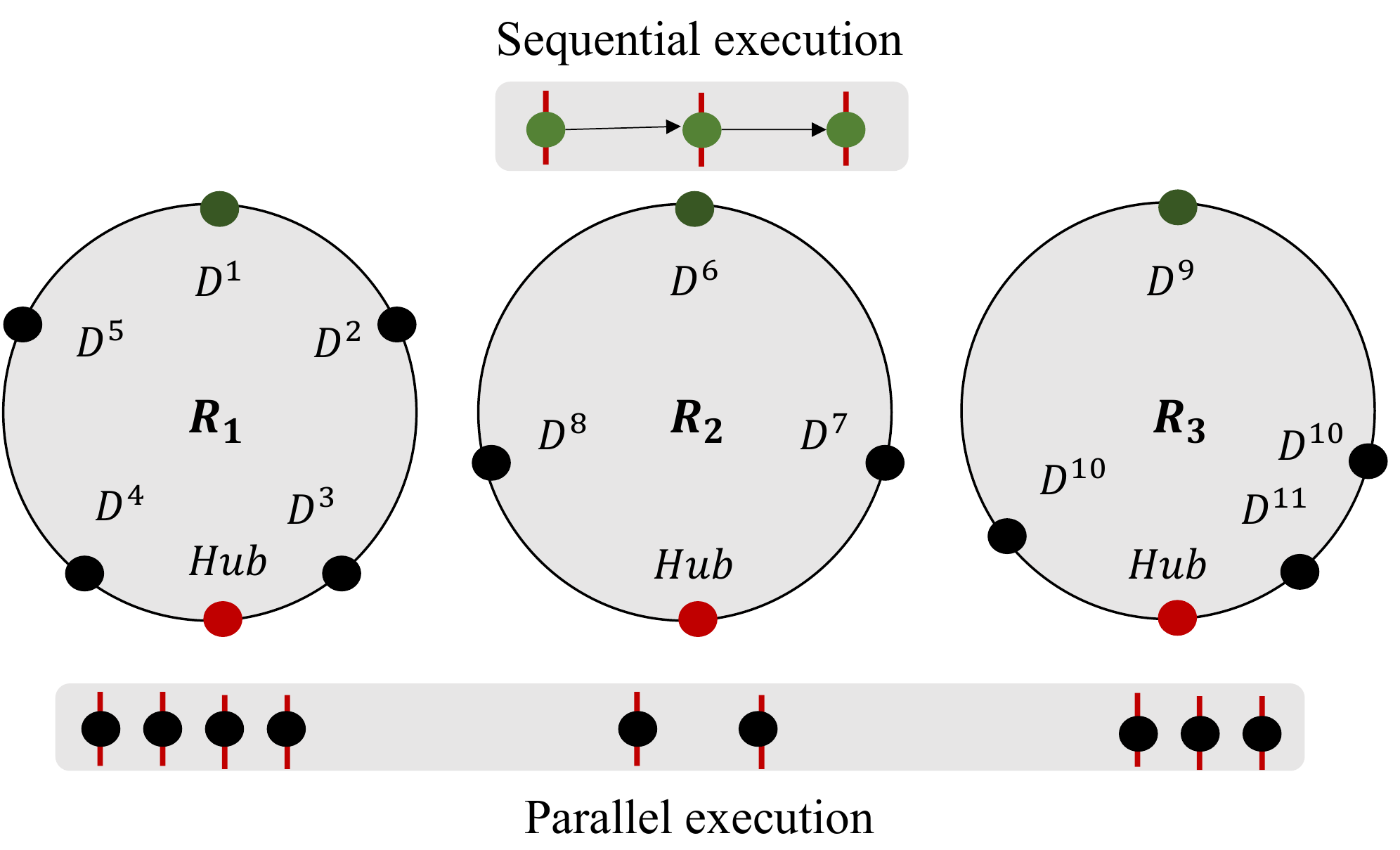}
	\end{center}
	%\vspace{0.2cm}
	%\end{framed}
\BBB
	\caption{Parallel rings}\label{fig:rin}
	\scriptsize{The figure shows parallel rings $R_1, R_2, R_3$ containing devices $D^1$ to $D^{12}$. Each ring has some incomparable devices shown in black, and a few comparable devices shown in green. Similarly, each ring has a red node that denotes the presence of a hub. There is only one shared hub for all $n$ devices in $m$ rings. The sequential execution bar on top shows the comparable devices during the peak load hour. The parallel execution bar below shows the incomparable devices during the non-peak load hour. Note that the figure above only shows a scalable construction of the original scheme with parallel rings. We do not show a failure resilient construction where instead of parallel rings parallel flowers can be actuated.}
	%\BBB\BBB
\end{figure}

The above-mentioned pre-profiling allows a different ordering and multiple dynamic rings' construction for the token circulation (as shown in Figure~\ref{fig:rin}). {Recall that in the static ring topology, each registered device strictly appears once to receive and forward the token. However, some devices, \textit{e}.\textit{g}., a garage opener and a smart car charger, receive fewer commands in a day, compared to other devices, \textit{e}.\textit{g}., smart switches, smart light bulbs, smart ovens, and door locks. Thus, in a static single ring, each device needs to participate equally in token transmission, whether they are participating in command execution or not. This problem of the static ring can be prevented by replacing the static ring construction with a dynamic ring.} The dynamic ring construction must be leveraged through the central hub and is based on two policies as below:

%However, in most cases, these devices just receive and forward the token and compute the puzzle to execute the command, not more than once a day. Therefore, the peak hour overhead of ordering $D^j$ in $\gamma_{D^i}$ can be optimized by replacing the static ring construction with a dynamic ring. The dynamic ring construction must be leveraged through the central hub and is based on two policies as below:

\smallskip
\noindent\textbf{[P1.]} {\em Based on device profiling, the hub must choose a device sub-set with a maximum number of incomparable devices to create a dynamic ring during the non-peak load hour.}

\noindent\textbf{[P2.]} {\em Based on device profiling, the hub must choose a device sub-set with a maximum number of comparable devices to create a dynamic ring during the peak load hour.}

%\smallskip
Assume that there are $N$ registered home devices that appear in a ring topology. Further, assume that half of these devices
are incomparable and receive fewer commands, each day. These commands are  sent to the device at any random interval after which a device starts working on the puzzle and, eventually, executes the command, regardless of the command execution by other devices, after finishing puzzle computation.  Now, assume that the remaining half of the devices are comparable and receive many commands, each day. For these comparable devices, an ordering is crucial regardless of the peak load hour(s). Thus, based on the profiling, an additional logic at the hub can dynamically create an additional ordering over the existing ordering (required by the user) to decide: which devices should be included in which of the rings? %Ideally, this dynamic device sub-set of size $j$ will be chosen such that there are no comparable devices in $r_i$, if it is a non-peak load hour, and, there are no incomparable devices in $r_i$. if it is a peak load hour.
%
%Thus, to enable a scalable yet sequential command execution during the peak load hour, a hub might initialize parallel rings. These parallel rings are meant to speed-up the command execution by including at most one comparable device in each ring and the remaining incomparable devices. In particular, at most one ring will be active during the peak load hour while all rings can be active simultaneously during the non-peak load hour.
%
%
%
Therefore, the profiling allows an optimal resource management such that: (\textit{i}) devices can execute the commands in parallel during certain non-peak load hours when resources are cheaper, and (\textit{ii}) devices execute the commands in a sequential manner during peak load-hours when resources are less cheap.

%\smallskip
\medskip
\noindent\textbf{Creation of parallel rings.} Initially, the hub knows an initial set of registered devices that are eligible to be included in any schedule. The hub maintains a device profile for each device. % and a load-profile. The device-profile $\varmathbb{P}$ for each of these devices defines the average load $\rho$, \textit{i}.\textit{e}., number of commands and peer-dependency $\gamma$, \textit{i}.\textit{e}., order with respect to peer devices. The load-profile is the average cost per hour and is changeable.
The hub creates at least two subsets of devices based on the device-profiles, one for the comparable devices and another for the incomparable devices. Next, the hub selects $x$ comparable devices, creates $x$ instances of empty rings, and assigns each of those comparable devices to one of the rings. These comparable devices are mutually exclusive and, therefore, during the peak load hour(s), only one of the rings needs to be active (hence, the corresponding comparable devices in that ring). Then, from another subset, all incomparable devices are selected and divided into either $x$ ring instances created in the previous step or in different numbers of rings. The rings for non-comparable devices may work in parallel with the ring for comparable devices, depending on the resources and workflow.

\subsection{Adaptation to Load-skew}
\label{sec:sly}
The proposed technique, which is based on continuous token rotation, can camouflage device communication and channel activity. The same token is used for downloading the commands from the hub at devices and to upload the data from devices to the hub as a result of command execution. {Based on the size of data, the devices request to upload, we can classify devices into two types: (\textit{i}) \textit{Skew devices} that wish to upload a large amount of data through the token, and hence, require more bandwidth.
A surveillance camera that needs to upload images periodically may be an example of skew devices. (\textit{ii}) \textit{Non-skew devices} that do not upload a large amount of data. Sensors such as ambient light sensors, smart meters, and smoke detectors that report numerical environmental data are examples of non-skew devices.} Recall that the latency of the proposed ring-based solution depends on two factors: (\textit{i}) the token length, which decides the time to transmit the token from one hop to next hop, and (\textit{ii}) the number of devices in a ring that indicates the number of hops the token needs to be forwarded.

Thus, when using only a single ring, the token size should be at least as big that can accommodate the maximum amount of the data produced by at least one skew device in one round of token. Of course, the token size  may be increased to accommodate the data produced by all devices in one round of token. Thus, placing all skew and non-skew devices in a single ring, the token length and latency will increase. In contrast, one may use a fixed small capacity single token in the ring, then the devices will require a faster circulation of the token in the ring.

In order to reduce latency, we can use parallel rings among skew and non-skew devices. In particular, we can place non-skew devices in one ring such that one round of a fixed (small)-sized token can satisfy their needs of data upload and place skew-devices in multiple rings.\footnote{{\scriptsize Of course, if placing all non-skew devices in a single ring increases latency, we can place them in multiple rings too.}}
Thus, skew devices do not incur latency overhead to non-skew devices. In all the rings, we can select different appropriate-sized tokens. Note that for the ring having skew devices, we can use the same-sized token, used for the ring having non-skew devices. However, as mentioned previously, in this case, the skew devices need more than one round of token circulation to upload the data.

%It is efficient to include parallel tokens, where each token might rotate at a different frequency to accommodate the bandwidth requirements.
Thus, in the case of skew and non-skew devices, the parallel ring-based solution can achieve efficiency; but it will reveal (\textit{i}) certain devices require a large-sized data upload request, or (\textit{ii}) high-frequency devices in the network, if we used the same-sized token in all rings and flow token selectively, as mentioned in Section~\ref{sec:paray}, unless non-skew devices require faster data upload request. Therefore, we can conclude that the concept of the parallel ring has a tradeoff between the efficiency and the leakage of device activity (within the workflow).

%\begin{figure}[h]??????>>>>>>>>>>>>>>>>>
\begin{wrapfigure}{r}{0.25\textwidth}	%\begin{framed}
\BBB\BBB\BB
	\begin{center}
		\includegraphics[scale=0.35]{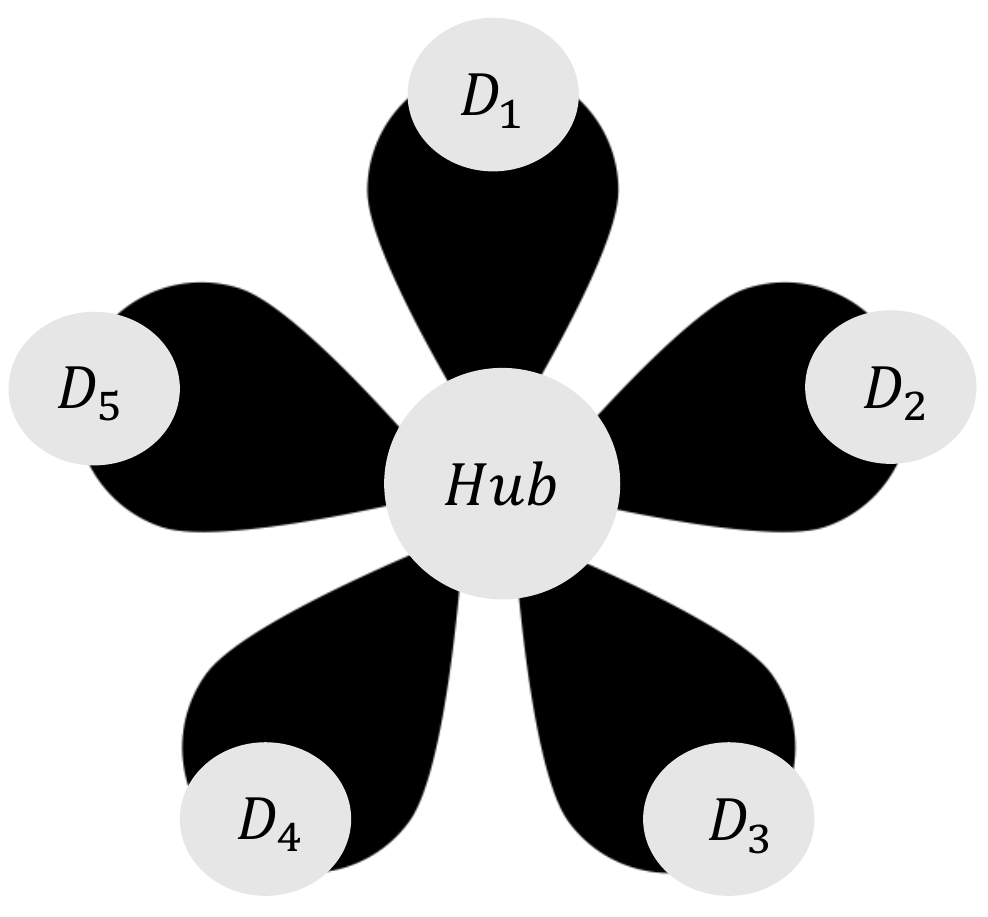}
	\end{center}
	%\vspace{0.2cm}
	%\end{framed}
\BBB
	\caption{Flower topology}
	\label{fig:flwr}
	\BBB\BBB
%\end{figure}
\end{wrapfigure}

\subsection{Resilience via Flower-based Ring}
\label{sec:flow}
The flower topology is the natural extension of ring topology as shown in Figure~\ref{fig:flwr}. The ring topology has a limitation that a single device failure might result in a broken ring and stop the token circulation. The adjacent peer devices in the ring formation might not realize this device failure and end up waiting for the token unpredictably. Therefore, we assume that a device failure (due to weak signal strength, mobility, or battery outage) can be detected within a bounded time interval in case the device does not respond. In particular, every time a device receives a token message directly from the central hub, the device must respond with an acknowledgment message back to the central hub. However, in our scheme (see Section~\ref{sec:decoup}), the central hub does not dispatch a dedicated token to each individual device. In fact, a common token is shared among all the devices within the ring. Therefore, the route of token circulation each time a device receives the token is directly through the central hub only, essentially forming a flower petal arch from hub-to-device and vice versa; as compared to the na\"{\i}ve ring where token circulates directly from one device to the next device. The communication overhead is twice the size of a ring in this flower-shaped ring topology. {Furthermore, the flower topology mixed with a ring network can deal with the complex network. For example, we may use the flower and ring topologies in a multi-story smart home, where a hub on a floor may form the flower topology containing devices situated on the floor and all hubs may participate a ring topology (assuming that hubs do not face failures). Thus, devices situated on different floors can communicate.}

Other concern is to deal with the {\em churn rate} where home devices might leave or join the network at any time. In the current model, it is non-trivial to decide for how long a sender device must wait for the response from a peer device. It is difficult to distinguish whether a device has left the network, or the device has not yet received the message due to the unreliable wireless medium. For example, devices can possibly move within the home periphery and outside the home periphery in which case devices might no longer receive `or' forward the token to other home devices. Since all devices are pre-registered and pre-profiled, devices can re-join the same group at a minimum authentication overhead. However, the frequent churning, \textit{i}.\textit{e}., specific devices leaving the network, reveals the activity status of these parallel rings and thereby the user activity.
{A na\"{\i}ve solution to prevent leakages from churning is to mimic the communication for a device, from the time it left the network to the time it re-joins the network. This could be achieved by, first, sending the token from the hub to the same device that just left the network; and then, mimicking the token-response by sending another packet. The token-response packet contains the MAC address of the device that left as the sender of the token-response packet and the MAC address of the hub as the receiver of the token-response packet. However, the wireless fingerprinting through signal strength -- that is outside the scope of this paper -- can potentially distinguish the actual communication and  mimicking the communication. }

\begin{wrapfigure}{r}{0.46\textwidth}
\BBB\BBB\BBB\BBB%\BBB\BBB\B
	\centering
	\includegraphics[scale=0.25]{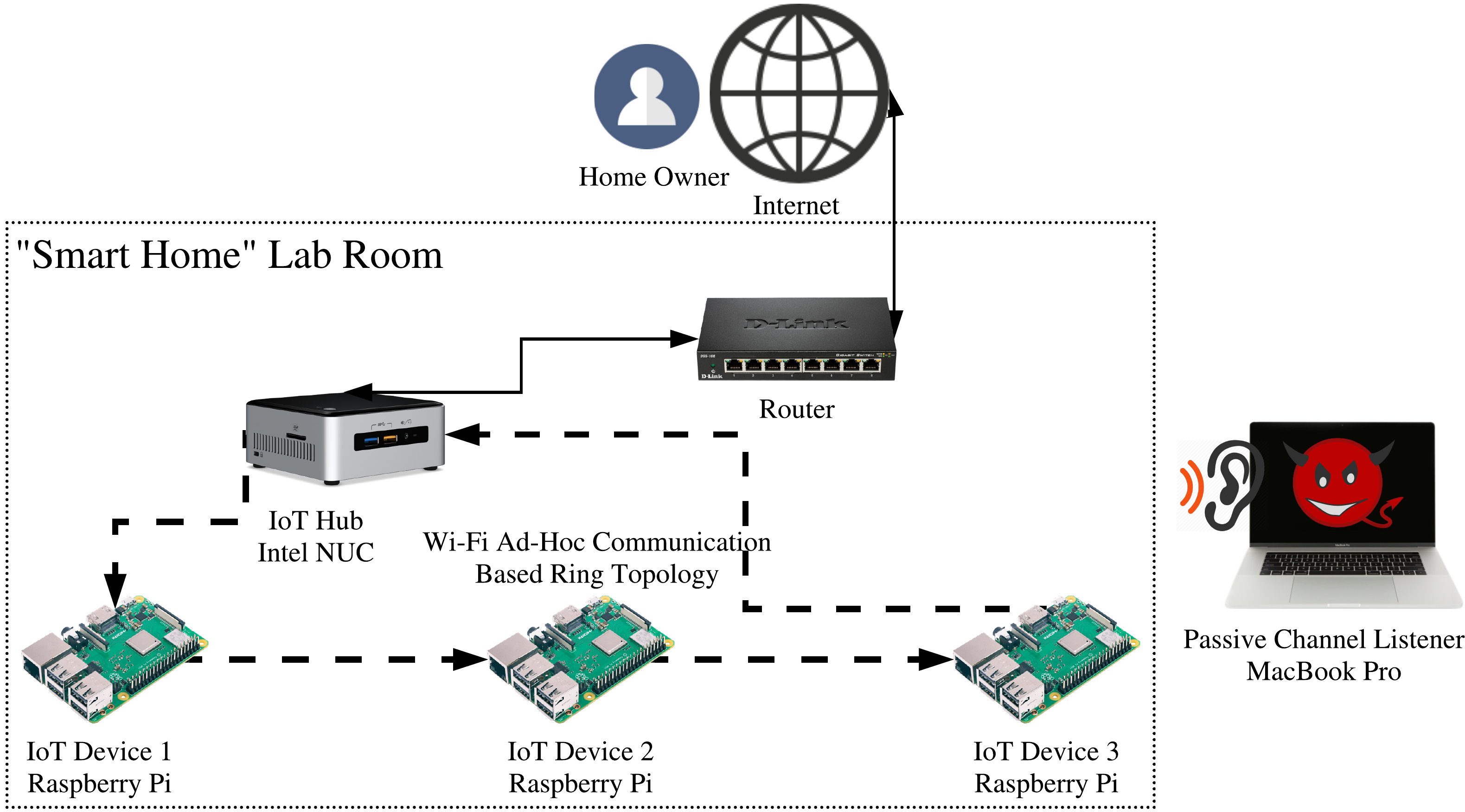}
	\caption{Experimental testbed in our lab}
	\label{fig:testbed}
\BBB\BBB\B
\end{wrapfigure}

\section{Experimental Evaluation}
\label{sec:Experimental Evaluation}
This section evaluates our proposed system using our prototype implementation. We describe the mock-up testing IoT application, experimental setup and overall results from our experiments.

\subsection{Experiment Setup}
%\B
In order to demonstrate the effectiveness and performance of our proposed architecture, we developed the prototype implementation in our lab (as shown in Figure~\ref{fig:testbed}). This proof-of-concept prototype implements the protocols described in Section~\ref{sec:decoup} and a test application with Python. In this mock-up IoT application, a device awaits and executes two types of commands, given by the homeowner. The ``Set'' command will change a variable in the program of the target device while the ``Read'' command will require the device to send the variable together with device status, \textit{e}.\textit{g}., RAM and CPU usages, back to the homeowner. This mock-up application is created to simulate two-way communication between the homeowner and the devices. Further, note that the application runs on top of the approach, we proposed in this paper.

Figure \ref{fig:testbed} depicts the architecture and configuration of the smart home testbed. An Intel NUC system is programmed to work as the hub $H$ that forwards information between the homeowner and smart devices. The hub $H$ equips with two network interfaces: (\textit{i}) Ethernet interface that has connections to receive the homeowner-defined schedule and sends the data of smart devices to the homeowner, and (\textit{ii}) WiFi interface that is used to communicate with smart home devices. Smart home device programs are deployed on three Raspberry Pis (${D^1, D^2, D^3}$) (3rd Gen B+ Model) that are equipped with built-in WiFi interfaces. WiFi interfaces on the hub $H$ and the devices ($D^1, D^2, D^3$) are configured to work in WiFi ad-hoc mode~\cite{Anastasihh}, which enables direct device-to-device communication. All WiFi interfaces are configured with pre-defined WiFi channels, static IP addresses, and routing information to have a ring topology. A MacBook Pro is deployed in a different room next to the testbed performs as an adversary, who listens and dumps all channel activities on the pre-defined channel into the pcap (packet capture) file for future analysis.

\subsection{Results}
\label{sec:res}
Based on the testbed described above, we performed different experiments to evaluate the proposed system and the approach. We first validate our system to check whether it could prevent the adversary from learning device activity from channel activity or not. Then, we explore the performance of the proposed ring topology communication with a set of experiments.

%\begin{figure}[!t]
%\BBB\BBB
%	\centering
%	\includegraphics[scale=0.25]{implementation.pdf}
%	\caption{Experimental testbed in our lab}
%	\label{fig:testbed}
%\BBB\B
%\end{figure}

\medskip
\noindent\textbf{Experiment 1: Decoupling channel activities from device activities.} To evaluate the effectiveness that our system protects against the passive channel listeners, we first defined a sequence of sample user commands, \textit{e}.\textit{g}., $D^1$: Set, $D^2$: Set, $D^2$: Read, and $D^3$: Read. In a one-minute experiment, these commands will be issued in 10 seconds intervals. We performed the experiment by executing the above-mentioned sequence in our proposed system and also over WiFi infrastructure network without a ring topology to compare with. The channel activities are recorded by the passive channel listener laptop deployed near our testbed.

\begin{wrapfigure}{r}{0.35\textwidth}
\BBB\BBB
    \centering
    \includegraphics[scale=0.4]{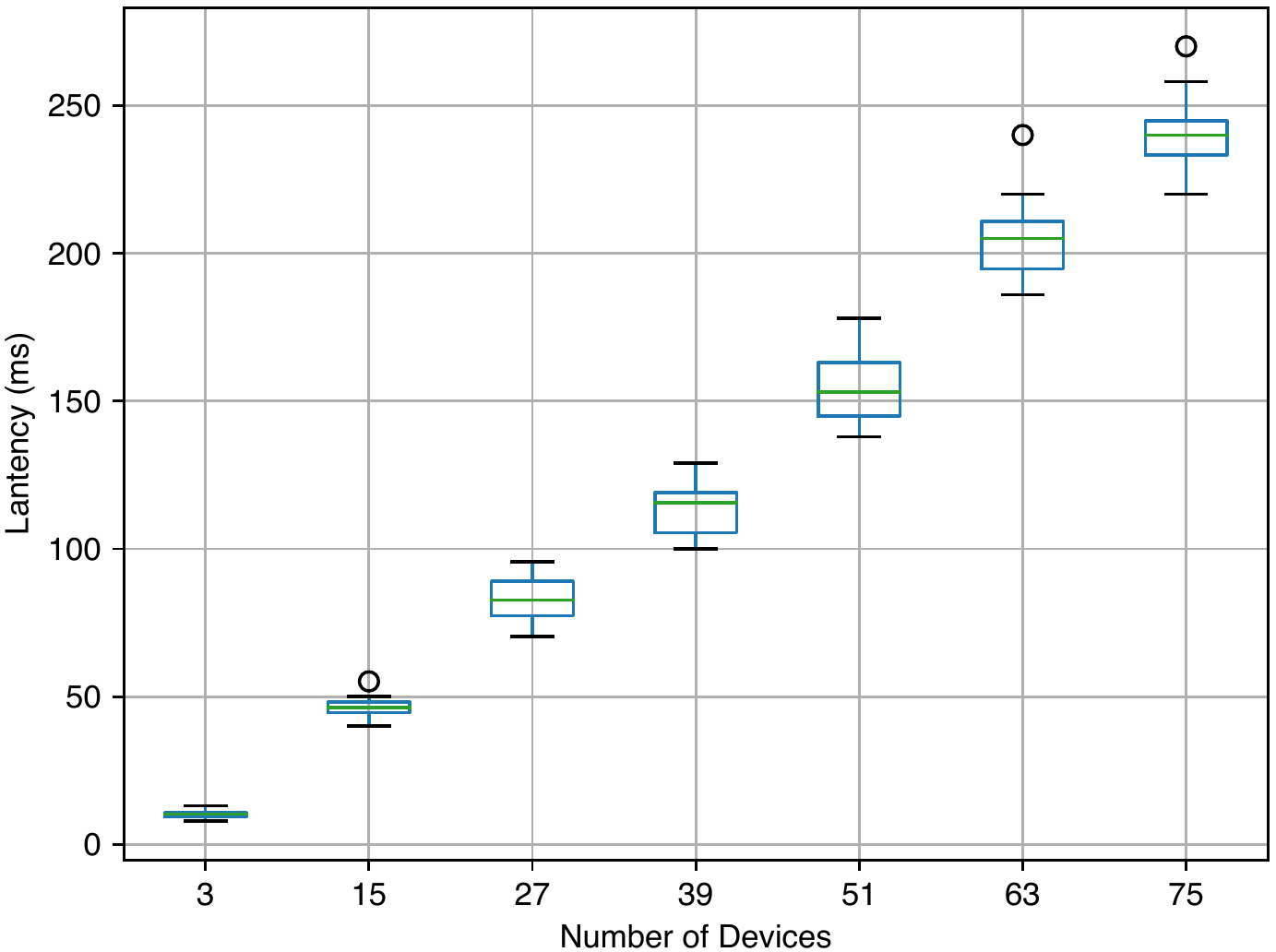}
\BB
    \caption{Experiment 2: Impact of ring topology on the latency}
    \label{fig:latency_exp}
    \BBB
\end{wrapfigure}

\begin{figure}[t]
\BBB\BBB\BBB
%\begin{center}
  \begin{minipage}[b]{.49\linewidth}
  \centering
  \includegraphics[scale=0.5]{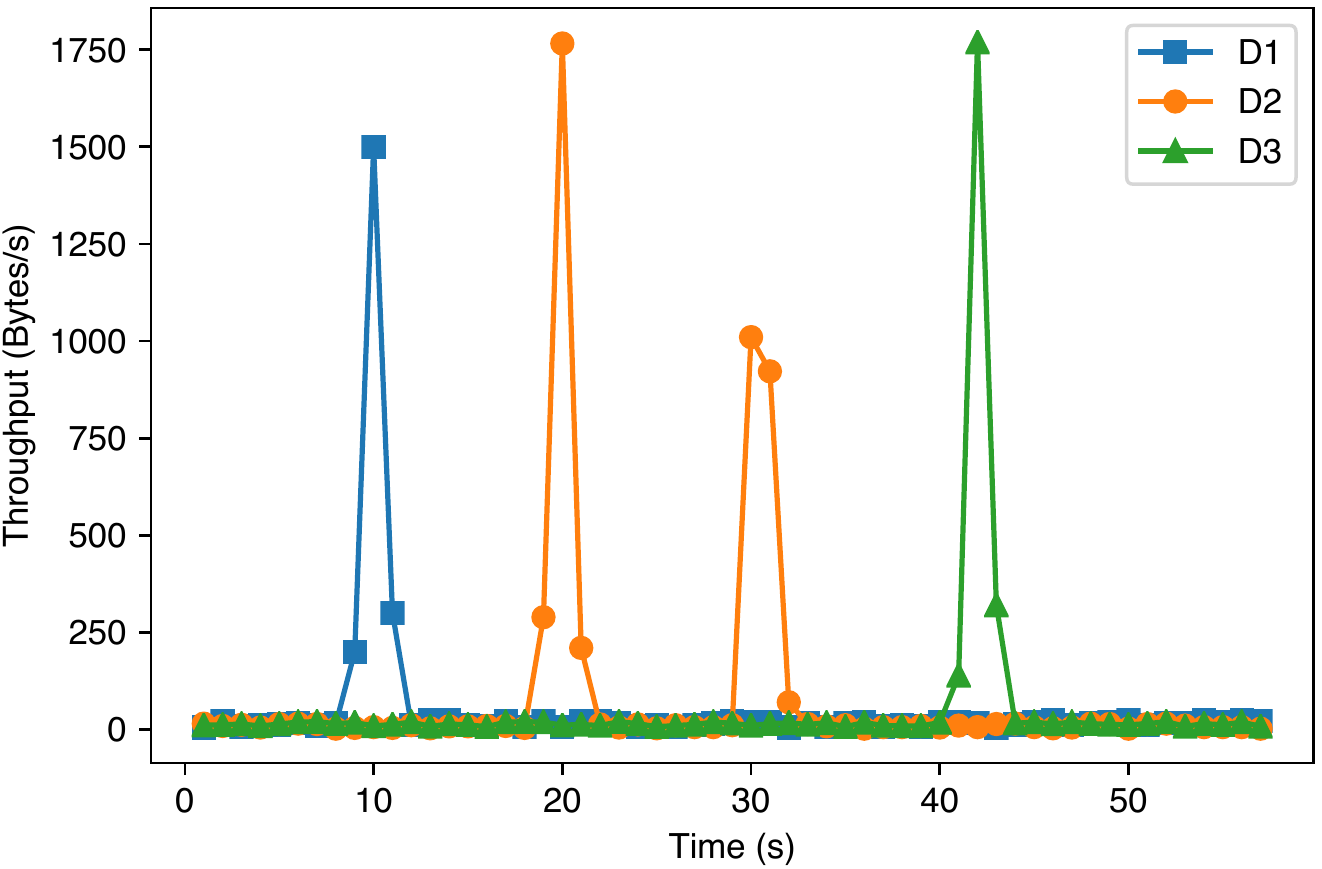}
  \subcaption{Devices working in common IoT settings.}
  \label{fig:infra}
  \end{minipage}
  \begin{minipage}[b]{.42\linewidth}
  \centering
    \includegraphics[scale=0.44]{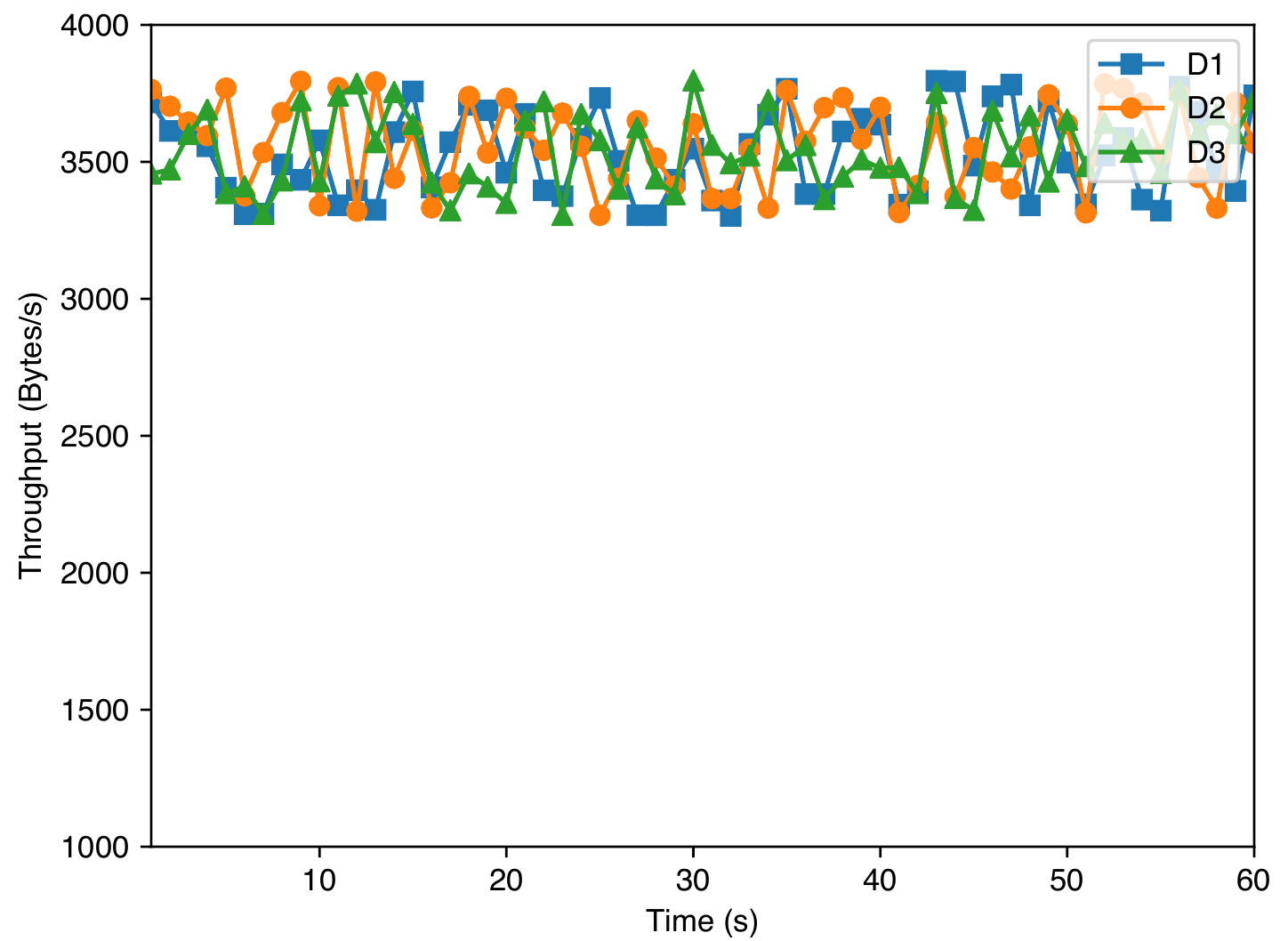}
    \subcaption{Devices working with our proposed system}
    \label{fig:adhoc}
  \end{minipage}
 %\end{center}
\BBB
\caption{Experiment 1: The adversarial view due to observing channel activities.}
\label{fig:channel}
\end{figure}

Figure \ref{fig:infra} shows the passive adversary's view, \textit{i}.\textit{e}., which device receives a message from the homeowner at which time due to channel activities in the experiment. It is clear that each time the user sends a command to devices or devices send data to the user, there will be a peak in the channel activity. Thus, the adversary infers the device activity and user-device interaction from channel activity. In contrast, Figure~\ref{fig:adhoc} shows the effectiveness of our proposed approach. Note that the channel activity patterns are completely eliminated, due to the token ring communication. %Now, the passive adversary cannot infer the device activity with the existence of our approach.

%\begin{figure}[!t]
%\BBB\BBB
%    \centering
%    \includegraphics[scale=0.4]{latency_exp.pdf}
%\BB
%    \caption{Experiment 2: Impact of ring topology on the latency}
%    \label{fig:latency_exp}
%\end{figure}

\medskip
\noindent\textbf{Experiment 2: Communication latency.} Instead of sending individual commands or data to/from devices/hub, the commands and data in our system are encapsulated in tokens and transmitted in a ring topology, which will incur additional communication latency. We performed experiments to evaluate the impact of increasing latency as the number of devices in the ring topology increases. Since we only have a very limited number of \emph{real} devices, we modified our protocol to simulate the scenario that includes a large number of devices to investigate an impact on communication latency. To achieve this goal, we add a counter in each token. When the hub generates the token, it sets the counter equals to the number of devices we want to simulate in the experiments. This counter is decreased by one when each device receives and forwards the token to the next device. When the last device ($D^3$ in our testbed) in the ring topology receives the token, it checks the value of the current counter. If the value of the counter is more than zero, $D^3$ forwards the token to the first device ($D^1$) to extend the ring topology. If the counter number is less than or equal to zero, the last device forwards the token back to the hub to complete a single round of the token. Here, since each device will receive the same token multiple times, we also need to prevent the device from solving same puzzles and executing same commands multiple times. To do so, a unique token ID is added to each token. Thus, the device solves the puzzle and executes the command only when the device gets a new token ID.

Figure~\ref{fig:latency_exp} shows that both mean and variation of the latency increase as more devices added into the ring topology. The mean of latency rises linearly at the beginning as each additional hop in the ring topology introduces more latency. After around 39 devices in the ring topology, the mean latency starts rising faster, since the length of the token also increases with the growth of a number of devices. Consequently, it may take more time at devices to transmit the token to the next hop. Of course, the ring latency is not affected, when there are a few devices, since the number of commands and toggle bit strings decrease in the token, as an decreasing number of devices. The variation of latency gets larger since it is more likely that the token transmission at more hops get delayed or re-transmitted because of unexpected interference or system lag at the devices.

\begin{table}[h]
\BBB
\begin{tabular}{|l|l|l|l|l|l|}
\hline
\textbf{\# Devices}       & { 3}       & { 27}      & { 51}   & { 63}   & {75}    \\ \hline
\textbf{ Avg. Token Len(bytes)} & {  1807}  & {  5024}  & {  8189} & {  9757} & {  11305} \\ \hline
\end{tabular}
\caption{Experiment 3: Average token length when having only one ring.}
\label{tbl:toklen}
\BBB\BBB
\end{table}

\begin{table}[h]
\BBB
\begin{tabular}{lllll}
\cline{1-4}
\multicolumn{1}{|r|}{\textbf{  State}}                   & \multicolumn{1}{c|}{  Idle}  & \multicolumn{1}{c|}{  IoT App w/o Ring Sys.} & \multicolumn{1}{c|}{  IoT App w/ Ring Sys.} &  \\ \cline{1-4}
\multicolumn{1}{|r|}{\textbf{  Avg. Power}} & \multicolumn{1}{c|}{  2.25W} & \multicolumn{1}{c|}{  3.03W}                 & \multicolumn{1}{c|}{  4.91W}                 &  \\ \cline{1-4}
                                                       &                           &                                           &                                           &  \\
                                                       &                           &                                           &                                           &

\end{tabular}
\BBB\BBB\B
\caption{Experiment 3: Average energy consumption of devices in different working states}
\label{tbl:energy}
\BBB%\BBB\BB
\end{table}

\medskip
\noindent\textbf{Experiment 3:  Communication and computation overheads.} It is also important to understand the additional overheads incurred due to our approach. We first measure how the length of token increases as the number of devices grows. Table~\ref{tbl:toklen} shows that the average token length grows linearly. We also use a USB power meter to measure the power consumption of the Raspberry Pi in different working states. Table \ref{tbl:energy} shows that our system introduces 63\% more power consumption to completely eliminate the channel activity patterns of all devices.

%experiments Set 1:

\medskip
\noindent
{\textit{Experiments related to extensions of the basic approach.} Below, we evaluate the impact of the parallel ring (given in Section~\ref{sec:extend}) on latency and token length. In particular, first, we examine how uniformly assigning devices into multiple parallel rings affects the performance (Experiment 4). Second, we perform experiments in the scenario where skew and non-skew devices exist all together (Experiment 5).}

\begin{figure}[!h]
	\BB%B\BBB%\BBB\BBB
	%\begin{center}
	\begin{minipage}[b]{.49\linewidth}
		\centering
		\includegraphics[scale=0.5]{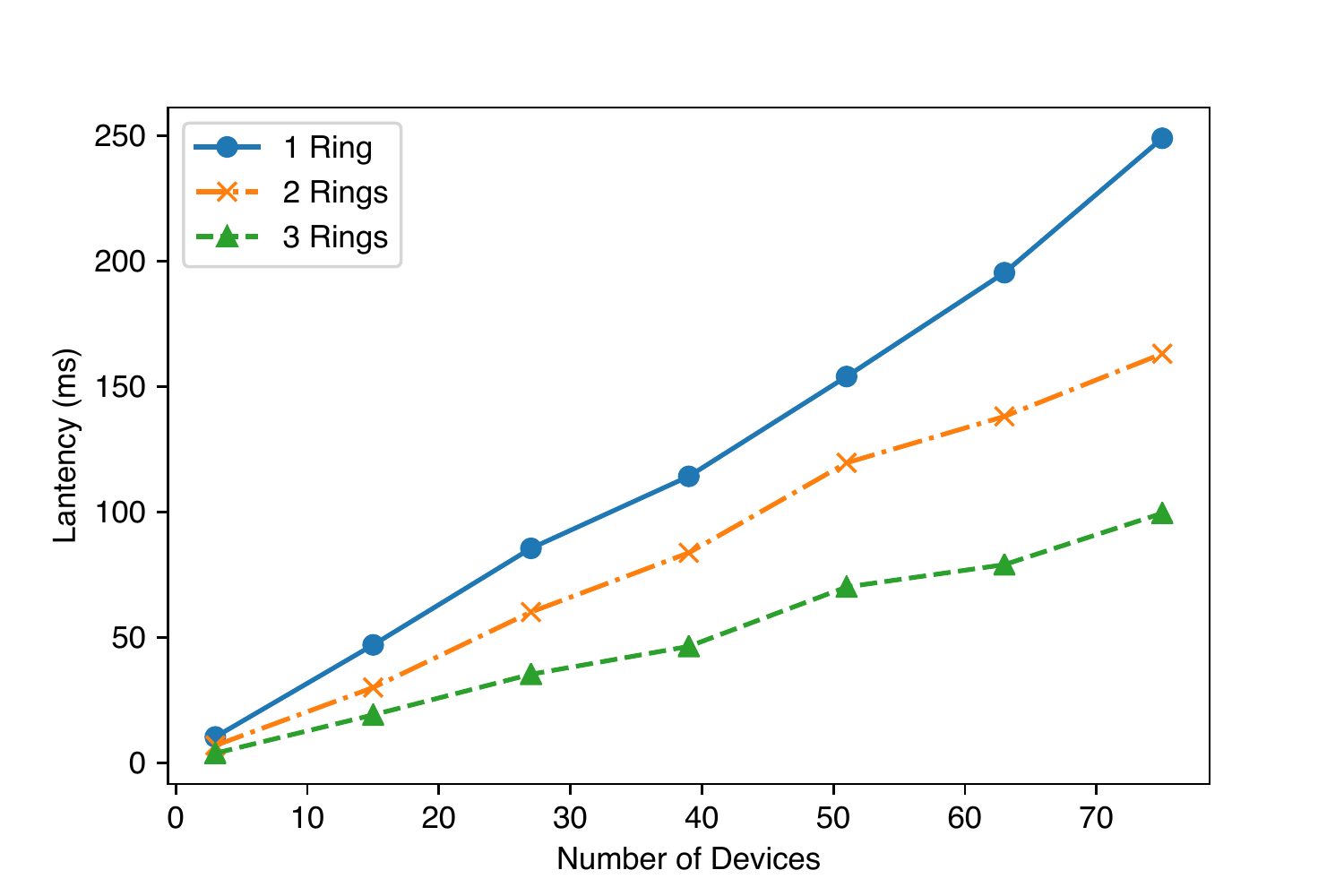}
		\subcaption{Impact of parallel rings on the latency.}
		\label{fig:mulrin_lat_exp}
	\end{minipage}
	\begin{minipage}[b]{.49\linewidth}
		\centering
		\includegraphics[scale=0.43]{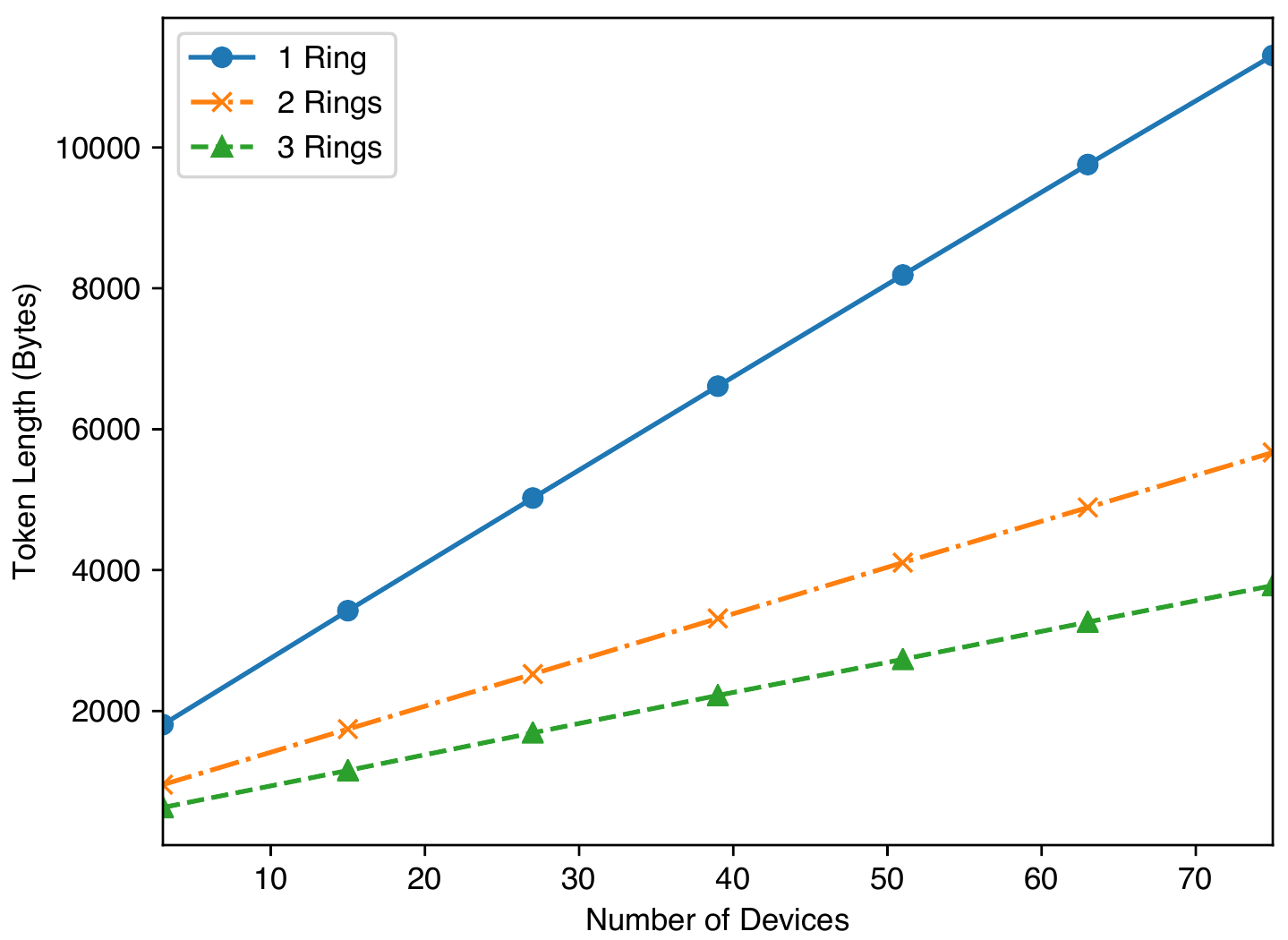}
		\subcaption{Impact of parallel rings on token length.}
		\label{fig:mulrin_tok_exp}
	\end{minipage}
	%\end{center}
	\BBB
	\caption{Experiment 4: Impact of non-skew devices in parallel rings.}
			\label{fig:nonskew}
\BBB
\end{figure}

\medskip
\noindent\textbf{Experiment 4: Scalability improvement through parallel rings.} This experiment evaluates the impact of allocating IoT devices into one, two, or three rings. IoT devices are assigned to one of the rings randomly, such that all rings contain an equal number of devices. We use six Raspberry Pis to simulate multiple IoT devices in different numbers of rings. In order to simulate a larger number of IoT devices in each ring, we use the same approach as described in Experiment 2. Figure~\ref{fig:mulrin_lat_exp} shows that the average latency of token transmission is lower in each ring when  multiple rings work in parallel. As the number of parallel rings increases, the number of IoT devices in each ring is reduced. Therefore, a token needs fewer hops in a ring before it comes back to the hub; consequently, it leads to lower latency. On the other hand, the less number of IoT devices in each ring also decreases the size of the token, as there are fewer data fields in the token in each ring. The reduced size of the token in multiple parallel rings can be seen in Figure~\ref{fig:mulrin_tok_exp}. {Note that in our experiment, the tokens were generated at the hub in a sequential manner and transmitted also in a sequential manner.}\footnote{{\scriptsize {The transmission of the token from the hub to the first devices in the rings can be done either in a sequential order by transmitting the $i^{\mathit{th}}$ token to the $i^{\mathit{th}}$ ring in the $i^{\mathit{th}}$ step or in parallel by sending tokens to all rings on different frequencies. The parallel token transmission requires that the hub should have multiple wireless interface cards, each of them works on a different channel, and each ring is allocated a different channel.}}}

{This experiment also shows the frequency of commands from a user to be delivered at devices. Observe that in a network having 75 non-skew devices, the user commands could be sent to devices around every 250 milliseconds when using only one ring, while using three rings, the user commands could be sent to devices around every 90 milliseconds.} As the results in Figure~\ref{fig:nonskew} show that distributing devices into parallel rings reduces communication overhead, as well as, communication latency for devices. However, increasing the number of parallel rings  leads to higher computation and communication overhead on the hub, because the hub has to communicate with multiple rings. In a scenario where a large amount of IoT devices exists, it is important to balance the IoT device communication latency and the hub communication overhead by tuning the number of rings that the hub manages.

\begin{figure}[!t]
 	\BBB\BBB\BBB\BBB
 	%\begin{center}
 	\begin{minipage}[b]{.49\linewidth}
 		\centering
 		\includegraphics[scale=0.5]{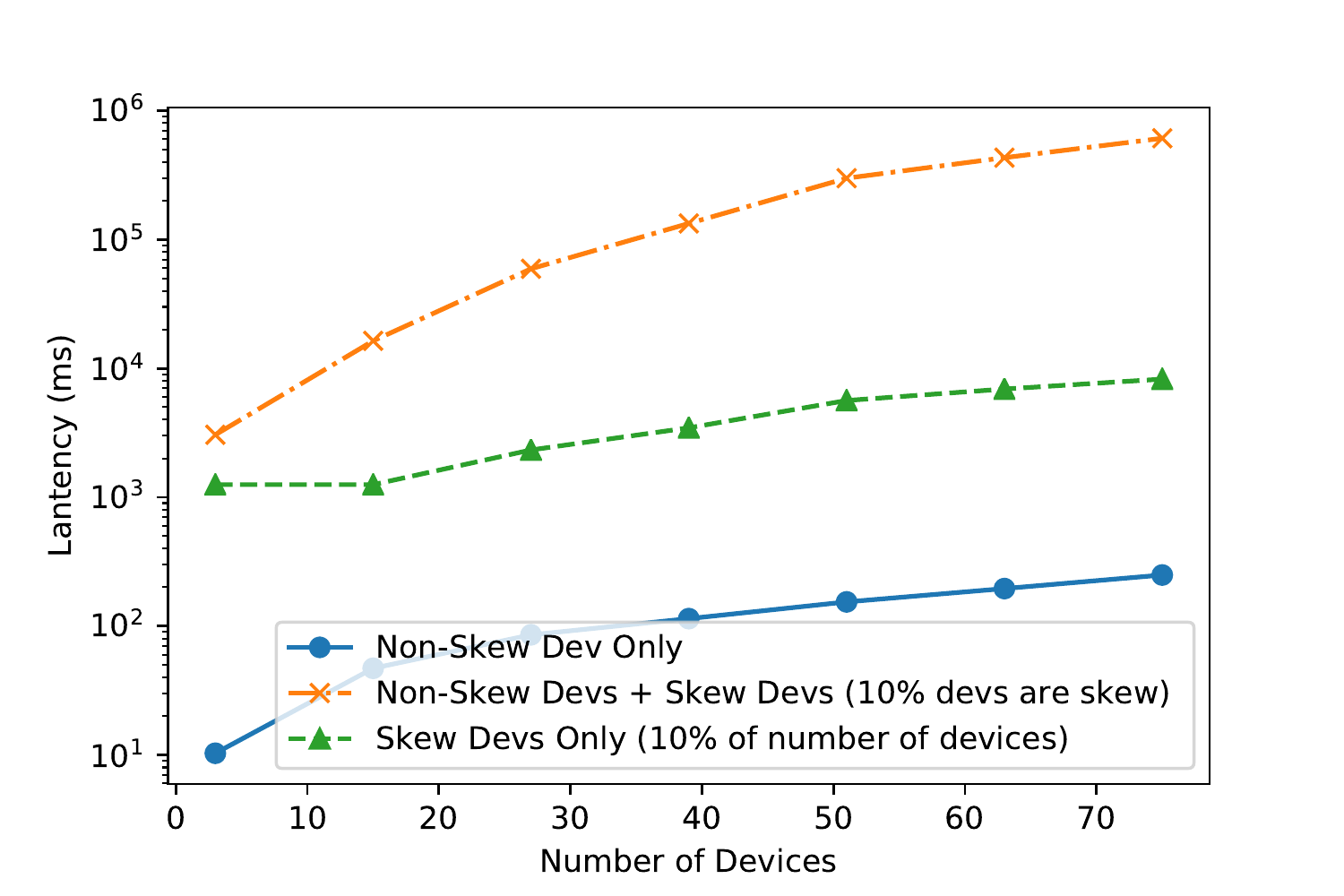}
 		\subcaption{Impact of parallel rings with skew devices on the latency}
 		\label{fig:skew_lat_exp}
 	\end{minipage}
 	\begin{minipage}[b]{.49\linewidth}
 		\centering
 		\includegraphics[scale=0.43]{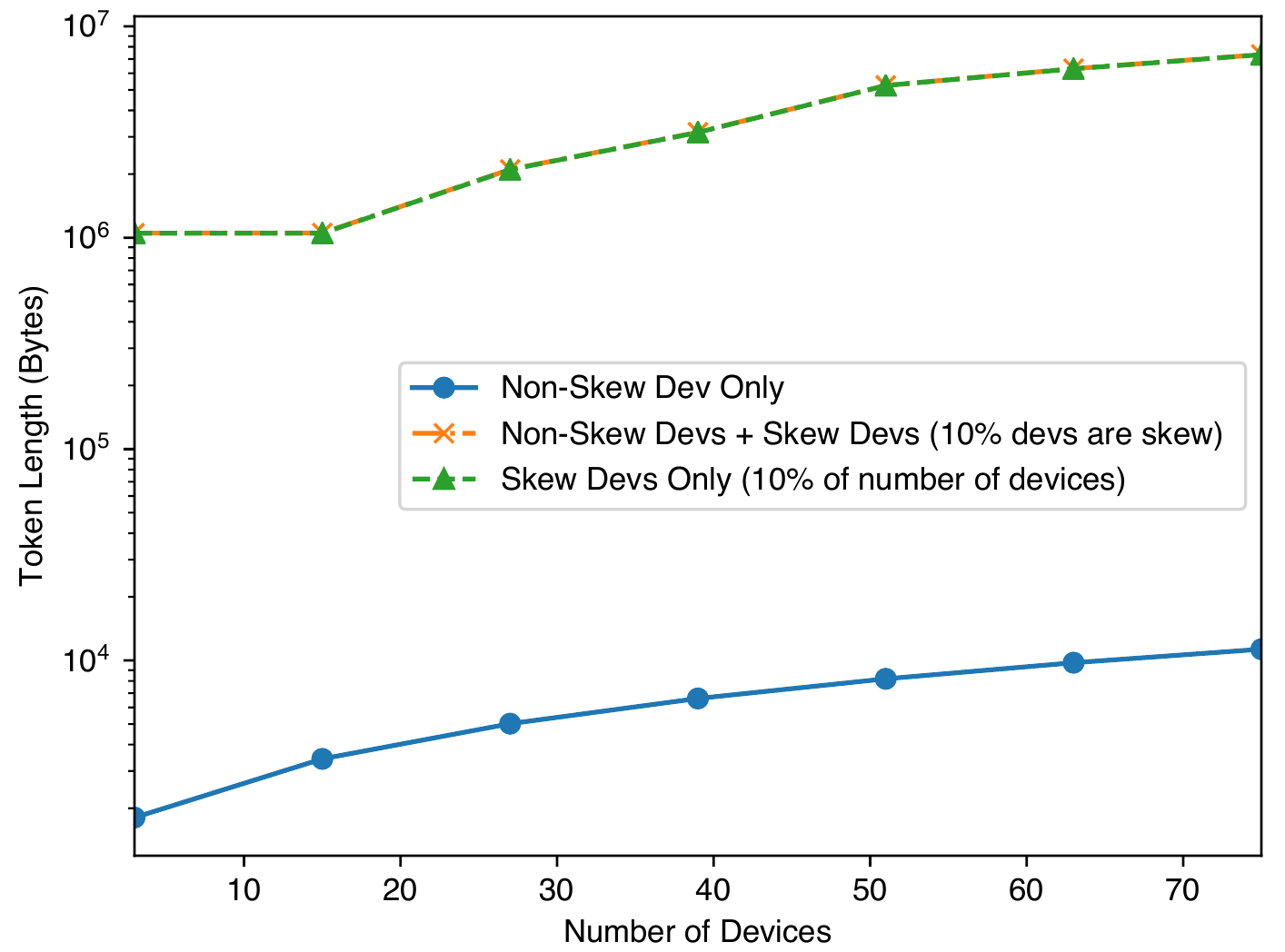}
 		\subcaption{Impact of parallel rings with skew devices on token length}
 		\label{fig:skew_tok_exp}
 	\end{minipage}
 	%\end{center}
 	\BBB
 	\caption{Experiment 5: Impact of skew devices in parallel rings.}
 	\label{fig:skew}
 %\B%BB
 \end{figure}

\medskip
\noindent\textbf{Experiment 5: Latency overhead for skew vs non-skew devices.} In the previous experiments, we assume that all IoT devices  require a similar size of data field in the token (\textit{i}.\textit{e}., non-skew devices). This experiment considers skew and non-skew devices. Different from non-skew devices that only receive and send small-sized data, skew devices usually require transmitting a much larger size data, \textit{e}.\textit{g}., an image or audio clip. In our experiment, we define the skew device as the device that requires a much larger size of data field (1MB) in the token, while non-skew devices only require 1KB of data field. We also assume that the number of skew devices is much less compared to the number of non-skew devices.

We perform experiments with a different number of IoT devices in two different settings: (\textit{i}) a ring that has all devices, and (\textit{ii}) two rings, one of them has only skew devices and another has only non-skew devices. {Note that in this experiment, we place all skew devices in one ring, and
it may increase the delay for skew devices. Of course, one may place skew (and non-skew) devices into multiple rings, as mentioned in Section~\ref{sec:sly}. However, here, our purpose is to show the practicality of parallel rings in the case of skew and non-skew devices.}

{In all experiments, we considered 90\% of devices as non-skew devices and 10\% of devices as skew devices. One can also vary the ratio between skew and non-skew devices; however, we fixed this ratio in this experiment, since the purpose of this experiment is to show that skew devices incur increased latency and token size.}

Figure~\ref{fig:skew_lat_exp} shows that, by assigning skew devices and non-skew devices into different rings, the latency of token transmission improves dramatically. In the setting where all devices are in the same ring, the token size is large due to the requirement of skew devices. Also, the number of hops that the token needs to be transmitted is high due to a large amount of non-skew devices in the same ring. Assigning skew devices and non-skew devices into different rings allows larger token only passing through a limited number of skew devices. Therefore, it reduces latency for both skew and non-skew devices. {Observe the first and the second data points (left two-points on the green-colored curve in Figure~\ref{fig:skew_lat_exp}), where we have less than 20 devices, we still need at least one skew device (since we assumed 10\% of all devices as skew devices). Thus, the latency of the first two data points of the skew devices curve is identical.} Figure~\ref{fig:skew_tok_exp} shows that separating skew devices and non-skew devices reduce the communication overhead of the non-skew devices, since the token size in the ring for non-skew devices is much smaller. The results in Figure~\ref{fig:skew} suggest that, when the variation of the required data field size from IoT devices is large, grouping and separating devices into different rings according to the required data field size could improve performance and reduce the device overhead. However, categorizing skew and non-skew devices may expose the device types to the adversary.

{From this experiment, it is clear now that as the ratio of skew and non-skew devices will increase (from what we assumed in this experiment), the token circulation latency and its size will increase, when allocating all devices in a single ring. Thus, this experiment shows the practicality of  parallel rings to deal with skew and non-skew devices.}

\section{Related Work}
\label{sec:rel}
There exist various IoT frameworks such as {\em Apple HomeKit}, {\em SmartThings}, {\em Azure IoT Suite}, {\em IBM Watson} IoT platform, {\em Brillo/Weave} platform by Google, {\em Calvin} IoT platform by Ericsson, {\em ARM mbed} IoT platform, {\em Kura} IoT project by Eclipse, interested readers may refer~\cite{doan,fram} for more details. %Overall, the smart home communication~\cite{tam,goto} is a network component that executes commands based on contextual factors.
{Characterizing and identifying IoT devices in a smart home or smart campus based on their network traffic has been explored recently.  These techniques are useful in various scenarios, such as IoT devices intrusion detection~\cite{DBLP:conf/iotdi/OrtizCL19,DBLP:conf/sp/AlrawiLAM19} and IoT devices management~\cite{DBLP:journals/tmc/SivanathanGLRWV19,traffic-iot-management2}.  However, there is also risks of device/user activity inference using the network traffic characteristics in a smart home. An adversary may identify the device/user-state or action by eavesdropping the traffic at the router/ISP level~\cite{DBLP:conf/sp/AlrawiLAM19,thir,pingpong} or sniffing the local wireless network channels~\cite{DBLP:journals/corr/abs-1808-02741,pingpong}.} To the best of our knowledge, none of the work (\textit{e}.\textit{g}.,~\cite{firs,thir,pingpong}) highlights the significance of secure device ordering in a smart home scenario. We highlight the presence of a pattern among smart home devices, such that a partial ordering on device activity is observed on a daily basis.~\cite{thir} presented a privacy-preserving traffic shaping scheme to mask the channel activity, thereby preventing the device or user activity at the ISP level. In~\cite{thir}, if the shaped traffic rate is lower than the device traffic. then the packets are queued, and if the shaped traffic rate is higher than the device traffic, then the dummy packets are added to cover the original traffic rate variations. However, these techniques do not avoid the inferences based on coupling device activities and channel activities. Similarly, a PingPong tool~\cite{pingpong} has been designed and evaluated to create a WAN sniffer that can detect the device activity. The sniffer taps into wireless signals and extracts packet-level signature, which can, further, be used to match and detect the device activity in the future communication stream.
Recent work~\cite{tifs,shen} provides a security framework for home devices to guarantee message anonymity and unlinkability, during the communication sessions from the hub to the device. The scheme is based on authentication and a one-time session key agreement in a three-way handshake protocol. However, as mentioned, authentication and encrypted messages cannot prevent the inferences from communication patterns.

\medskip
\noindent\textbf{IoT frameworks.} There exist a number of IoT frameworks~\cite{doan,fram} based on a general\footnote{{\scriptsize The existing literature and industrial IoT home frameworks consider a device-to-hub network connectivity model, \textit{e}.\textit{g}., CloudCam, Philips SmartLight, and assume a trustworthy hub.}} model that includes IoT devices, a backend cloud, and a proxy hub. These frameworks support a variety of IoT devices from many vendors (\textit{e}.\textit{g}., Amazon, Samsung, Google, Philips Hue, Nest, Belkin) that require capabilities to encrypt/decrypt messages and store the keys. However, the above-mentioned device security model requires capabilities to encrypt/decrypt messages and store the keys. Thus, on both (communication and security) fronts, currently available IoT devices are sufficient to implement our scheme. {Also, existing IoT frameworks already support the scheduling function that is required to implement workflow.} Here, below, we provide a brief overview of these IoT frameworks.

\medskip
\noindent\textbf{Apple HomeKit:}
This IoT framework is dedicated to smart home device connectivity. It leverages the connectivity for IoT home appliances and accessories through smartphone iOS apps. The iOS app {\em Home} allows the devices to join/leave the home network, customize, and control the home environment. In the HomeKit architecture, IoT devices connect to the platform either directly or through proxy gateways that supports ZigBee and Z-Wave communication protocols. However, the IoT devices that directly connect through {\em HomeKit} accessory protocol can communicate through LAN, WiFi or BLE, instead of ZigBee and Z-wave  protocols. For example, tvOS 10 supports the {\em Homekit} framework and acts as the hub for IoT home devices. The security layer in {\em Homekit} includes {\em Perfect Forward Secrecy} (PFS) and secure communication over Transport Layer Security (TLS) or Datagram TLS with AES128-GCM, AES256 and SHA256. It must be noted that an owner can choose actions for IoT devices through Siri service in Homekit. However, it is not possible to schedule a hub-dependent sequence of device actuation. The PFS ensures that any future communication is secure, and the leakage of long-term keys in the future cannot reveal the sessions from the past. In addition, the applications' access to home data is based on the permission model, and the iOS system data is secure against memory-based attacks through {\em Address Space Layout Randomization} (ASLR) technique. {In HomeKit framework, the workflow is instantiated by creating {\em automation} ~\cite{ref-iot-wf-apple}. In {\em automation}, the set of actions could be triggered by either a device/people-related event or at a certain time. The defined {\em automations} are instantiated by a home hub device.}

\medskip
\noindent\textbf{Amazon Web Service (AWS):} This IoT platform provides a ubiquitous  connectivity between the IoT devices and the AWS cloud. The AWS architecture includes: (\textit{a}) {\em device gateway} providing connectivity among IoT devices and the cloud services through MQTT (Message Queue Telemetry Transport), SSL (Secure Socket Layer), TLS, Websockets and HTTP (Hyper Text Transfer Protocol) 1.1; (\textit{b}) {\em device shadows} maintaining a virtual replica of the original device also keeps synchronizing the device state. In case a device is offline the device shadow retains the last visible state of the device and all pending upgrades can be restored once the device is online; (\textit{c}) {\em rule engine} providing a policy execution on the published data and transforming it into subscriber-appropriate format; (\textit{d}) {\em registry} maintaining the meta-level information (\textit{e}.\textit{g}., device name, identity, vendor, other attributes, etc) about connected devices. The security architecture of AWS includes authentication based on $X.509$ certificates, confidentiality through SSL/TLS based secure key exchange, access control through policy specification, forward secrecy through TLS cipher suites such as AES128-GCM-SHA256, ECDHE-ECDSA-AES128-GCM-SHA256, and, AES256-GCM-SHA384. {AWS provides rich functionalities to support the workflow instantiation in IoT applications. For example. AWS IoT jobs can be used to define a set of remote operations that are sent to and executed on one or more devices connected to AWS IoT. Users could define the timings of an AWS IoT job to schedule the job execution~\cite{ref-iot-wf-aws-job}.}

\medskip
\noindent\textbf{Samsung SmartThings:} This IoT platform is dedicated to smart home environments and appliance connectivity through mobile phone apps. The {\em SmartThings} framework is composed of a cloud backend, hub, mobile client applications, and  IoT devices. In {\em SmartThings} framework, the hub interacts with the home devices and the cloud services. The hub provides connectivity through several communication protocols such as ZigBee, Z-Wave, WiFi, and BLE. In addition, the cloud-connected devices utilize {\em OAuth/OAuth2} protocol for authentication and SSL/TLS for message transmission. In addition, the hub supports AES-128 bit encryption for all communication with ZigBee and Z-Wave enabled products. {The workflow could be implemented in the SmartThings framework by schedule future executions using the Scheduling API~\cite{ref-iot-wf-samsung}. Executions could either be done exactly once at the specified day and time or in a repetitive manner that is defined by Cron expressions.}

\medskip
\noindent\textbf{Azure IoT Suite:} The suite includes IoT devices, cloud services, and the hub to provide secure connectivity. The cloud is entitled to send commands and notifications for the IoT devices through the hub. In this IoT platform, the hub supports MQTT and HTTP protocols to enable this bi-directional connectivity. The security layer provides device authentication, access control, and communication security. The device authentication is based on HMAC-SHA256 signed token along with the unique device identity. The access control and authorization are based on permission policies defined in the {\em Azure Active Directory}. The SSL/TLS protocol is used for a secure handshake, mutual authentication, and session secrecy. There exist various other frameworks such as {\em IBM Watson} IoT platform, {\em Brillo/Weave} platform by Google, {\em Calvin} IoT platform by Ericsson, {\em ARM mbed} IoT platform, {\em Kura} IoT project by Eclipse, interested readers may refer~\cite{fram} for more details. {Similar to AWS IoT, Azure IoT Hub allows a user to schedule and track the progress of any activity on a set of devices~\cite{ref-iot-wf-azure-job}.}

\medskip
\noindent
{\textbf{IFTTT:} For the case that some IoT frameworks and applications do not have workflow properties natively, or there are multiple different IoT frameworks at home, a user could implement the workflow functionality by using a third-party web service that supports trigger-action style programming. The most popular platform is {\em IFTTT} ({\em If This, Then That}), which allows users to define simple conditional statements as {\em applets}. An {\em applet} receives trigger information from supported services and performs predefined actions on other services. For example, a user could define an applet to turn on the light bulb controlled by Apple {\em HomeKit}, when the {\em SmartThings} ambient light sensor detects the dark environment. {\em IFTTT} is widely supported by a large number of services and applications. Till early 2018, there are more than 600 services connect to {\em IFTTT}~\cite{ref-iot-wf-ifttt}.}

\section{Conclusion}
{This paper focuses on the security and privacy challenges in smart homes due to the execution of different types of workflows. In particular, this paper deals with the problem of inferring the user activity -- which may, in turn, reveal the user privacy -- by coupling the device activities (due to workflow execution) and corresponding network traffic analysis. The paper proposes solutions based on a cryptographically secure token circulation in a ring network of devices to prevent revealing any device activity. In order to deal with a large number of devices, device failure, different complex networks, different types of devices based on the data they generate, we provide solutions based on either multiple rings that work in parallel or using a flower topology. Our experimental results show the practicality of the solution and evaluate the techniques' performance in terms of latency and overheads.}

\medskip
\noindent
{\textbf{Future directions.} The algorithms proposed in this paper scales well for smart home devices. An interesting direction should be to implement the algorithm at a much larger scale in different scenarios, for example, a smart enterprise building and a smart industrial workplace. Another interesting direction would be to design an efficient method for existing hub devices to dynamically create device profiles. It needs to learn the behavior of all home devices by the hub (and then implement the parallel rings). One direction may be designing and implementing methods to prevent attacks based on wireless fingerprinting-based attacks. Such methods will be useful to prevent an adversary from learning due to wireless fingerprinting-based attacks, when a device leaves the network and our approach sends phantom messages to ghost devices. An important direction is to design a secure algorithm, while also preserving the device activities, in the case of multicast streaming applications that require transmitting a large amount of data to multiple endpoints in realtime with strict synchronization constraints.}

%\bibliographystyle{abbrv}
%{\scriptsize \bibliography{RelatedWorkCamera}}
%\bibliography{RelatedWorkGuoxiAdded}

\begin{thebibliography}{10}

\bibitem{old-people}
\url{https://www.prb.org/aging-unitedstates-fact-sheet/}.

\bibitem{55old}
\url{https://www.mylifesite.net/blog/post/pros-cons-55-active-adult-communities/}.

\bibitem{guoxi1}
In {\em Insight' into Home Automation Reveals Vulnerability in Simple IoT
  Product, available at URL:
  \url{https://securingtomorrow.mcafee.com/mcafee-labs/insight-into-home-automation-reveals-vulnerability-in-simple-iot-product
  }}.

\bibitem{2guoxi}
In {\em IEEE OUI (Organizationally Unique Identifier)}.

\bibitem{ref-iot-wf-samsung}
Smartthings developers.
\newblock
  \url{https://smartthings.developer.samsung.com/docs/smartapps/scheduling.html}.

\bibitem{ref-iot-wf-apple}
Create home automations with the home app.
\newblock {\em Apple Support}, Sep 2019.
\newblock \url{https://support.apple.com/en-us/HT208940}.

\bibitem{DBLP:journals/corr/abs-1808-02741}
A.~Acar et~al.
\newblock Peek-a-boo: {I} see your smart home activities, even encrypted!
\newblock {\em CoRR}, abs/1808.02741, 2018.

\bibitem{DBLP:journals/suscom/AlhassounUV19}
N.~S. Alhassoun and other.
\newblock Context-aware energy optimization for perpetual iot-based safe
  communities.
\newblock {\em {SUSCOM}}, 22:96--106, 2019.

\bibitem{DBLP:conf/sp/AlrawiLAM19}
O.~Alrawi et~al.
\newblock {SoK}: Security evaluation of home-based {IoT} deployments.
\newblock In {\em {IEEE SP}}, pages 1362--1380, 2019.

\bibitem{fram}
M.~Ammar et~al.
\newblock Internet of things: A survey on the security of {IoT} frameworks.
\newblock {\em {JISA}}, 38:8 -- 27, 2018.

\bibitem{Anastasihh}
G.~Anastasi et~al.
\newblock {Wi-fi in ad hoc mode: a measurement study}.
\newblock In {\em {PerCom}}, pages 145--154, 2004.

\bibitem{firs}
N.~Apthorpe et~al.
\newblock Closing the blinds: Four strategies for protecting smart home privacy
  from network observers.
\newblock {\em CoRR}, abs/1705.06809, 2017.

\bibitem{thir}
N.~Apthorpe et~al.
\newblock Spying on the smart home: Privacy attacks and defenses on encrypted
  {IoT} traffic.
\newblock {\em CoRR}, abs/1708.05044, 2017.

\bibitem{carnemolla2018ageing}
P.~Carnemolla.
\newblock Ageing in place and the internet of things--how smart home
  technologies, the built environment and caregiving intersect.
\newblock {\em Visualization in Engineering}, 6(1):7, 2018.

\bibitem{doan}
T.~T. Doan et~al.
\newblock Towards a resilient smart home.
\newblock In {\em {IoT S\&P}}, pages 15--21, 2018.

\bibitem{DBLP:conf/sigcomm/HamzaGS18}
A.~Hamza et~al.
\newblock Combining {MUD} policies with {SDN} for iot intrusion detection.
\newblock In {\em {IoT S{\&}P}}, pages 1--7, 2018.

\bibitem{ref-iot-wf-ifttt}
K.~Johnson.
\newblock Ifttt raises \$24 million, led by salesforce ventures.
\newblock {\em VentureBeat}, Apr 2018.
\newblock
  \url{https://venturebeat.com/2018/04/26/ifttt-raises-24-million-led-by-salesforce-ventures/}.

\bibitem{tifs}
P.~Kumar et~al.
\newblock Anonymous secure framework in connected smart home environments.
\newblock {\em {IEEE} Trans. Information Forensics and Security},
  12(4):968--979, 2017.

\bibitem{DBLP:journals/popets/MartinMDFBRRB17}
J.~Martin et~al.
\newblock A study of {MAC} address randomization in mobile devices and when it
  fails.
\newblock {\em PoPETs}, 2017(4):365--383, 2017.

\bibitem{DBLP:conf/iotdi/OrtizCL19}
J.~Ortiz et~al.
\newblock {DeviceMien}: network device behavior modeling for identifying
  unknown {IoT} devices.
\newblock In {\em {IoTDI}}, pages 106--117, 2019.

\bibitem{DBLP:conf/codaspy/Panwar0WMV19}
N.~Panwar et~al.
\newblock Verifiable round-robin scheme for smart homes.
\newblock In {\em Proceedings of the Ninth {ACM} Conference on Data and
  Application Security and Privacy, {CODASPY}}, pages 49--60, 2019.

\bibitem{rlock}
R.~L. Rivest et~al.
\newblock Time-lock puzzles and timed-release crypto.
\newblock Technical report, {MIT/LCS/TR-684, MIT} {L}ab for {C}omputer
  {S}cience, 1996.

\bibitem{ref-iot-wf-azure-job}
Robinsh.
\newblock Understand azure iot hub jobs.
\newblock {\em Understand Azure IoT Hub jobs, Microsoft Docs}.
\newblock
  \url{https://docs.microsoft.com/en-us/azure/iot-hub/iot-hub-devguide-jobs}.

\bibitem{shen}
J.~Shen et~al.
\newblock Secure data uploading scheme for a smart home system.
\newblock {\em Information Sciences}, 453:186 -- 197, 2018.

\bibitem{DBLP:journals/tr/SiboniSMBMBSE19}
S.~Siboni et~al.
\newblock Security testbed for internet-of-things devices.
\newblock {\em {IEEE} Trans. Reliability}, 68(1):23--44, 2019.

\bibitem{traffic-iot-management2}
A.~Sivanathan et~al.
\newblock Characterizing and classifying {IoT} traffic in smart cities and
  campuses.
\newblock In {\em {INFOCOM WKSHPS}}, pages 559--564, 2017.

\bibitem{DBLP:journals/tmc/SivanathanGLRWV19}
A.~Sivanathan et~al.
\newblock Classifying iot devices in smart environments using network traffic
  characteristics.
\newblock {\em {IEEE} Trans. Mob. Comput.}, 18(8):1745--1759, 2019.

\bibitem{stites2014user}
D.~Stites and K.~Skinner.
\newblock User privacy on {iOS} and {OS X}, 2014.
\newblock The Apple Worldwide Developers Conference.

\bibitem{pingpong}
R.~Trimananda et~al.
\newblock {PingPong}: Packet-level signatures for smart home device events.
\newblock {\em CoRR}, abs/1907.11797, 2019.

\bibitem{DBLP:conf/smartcomp/UddinNBWZHACSHD16}
M.~Y.~S. Uddin et~al.
\newblock The {Scale2} multi-network architecture for {IoT}-based resilient
  communities.
\newblock In {\em {SMARTCOMP}}, pages 1--8, 2016.

\bibitem{DBLP:conf/ccs/VanhoefMCCP16}
M.~Vanhoef et~al.
\newblock Why {MAC} address randomization is not enough: An analysis of wi-fi
  network discovery mechanisms.
\newblock In {\em {AsiaCCS}}, pages 413--424, 2016.

\bibitem{ref-iot-wf-aws-job}
P.~Waher et~al.
\newblock Using the aws iot jobs apis.
\newblock {\em Amazon}, 2016.
\newblock
  \url{https://docs.aws.amazon.com/iot/latest/developerguide/jobs-api.html}.

\bibitem{watson}
G.~J. Watson et~al.
\newblock Lost: Location based storage.
\newblock In {\em ACM Workshop on Cloud Computing Security}, pages 59--70,
  2012.

\bibitem{DBLP:journals/tecs/ZhouSY19}
L.~Zhou et~al.
\newblock A lightweight cryptographic protocol with certificateless signature
  for the internet of things.
\newblock {\em {ACM} Trans. Embedded Comput. Syst.}, 18(3):28:1--28:10, 2019.

\bibitem{zhu2017embedded}
Y.~Zhu.
\newblock {\em Embedded systems with arm cortex-m microcontrollers in assembly
  language and c}.
\newblock E-Man Press Llc, 2017.

\end{thebibliography}
\end{document}